\documentclass[pre,preprint,showpacs,showkeys]{revtex4}
\usepackage{graphicx}
\usepackage{subfigure}

\begin{document}

\title{Monte Carlo simulation study of diblock copolymer self assembly}

\author{George J. Papakonstantopoulos$^1$\footnote{Author to whom correspondence should be addressed. Electronic mail:gjpapakonsta@gmail.com}}
\author{Kostas Ch. Daoulas$^2$}
\author{Marcus M\"uller$^2$}
\author{Juan J. de Pablo$^3$}
\affiliation{$^1$Department of Chemical Engineering,University of Wisconsin-Madison, Madison, WI 53706 - Currently: Department of Materials Science at Goodyear Tire \& Rubber Company, Akron, OH, 44305}
\affiliation{$^2$Institut f\"{u}r Theoretische Physik, Georg-August Universit\"{a}t,37077 G\"{o}ttingen, Germany and Department of Physics, University of Wisconsin-Madison,Madison, Wisconsin 53706, USA - Currently: Department of Polymer Theory, Max Planck Institute for Polymer Research, Mainz, P.O. Box 3148}
\affiliation{$^3$Institut f\"{u}r Theoretische Physik, Georg-August Universit\"{a}t,37077 G\"{o}ttingen, Germany and Department of Physics, University of Wisconsin-Madison,Madison, Wisconsin 53706, USA}
\affiliation{$^4$Department of Chemical Engineering,University of Wisconsin-Madison, Madison, WI 53706 - Currently: Institute of Molecular Engineering, University of Chicago, Chicago, IL, 60637}

\date{\today}

\begin{abstract}
A technique is presented which maps the parameters of a bead spring model, using the Flory 
Huggins theory, to a specific experimental system. By keeping only necessary details, for 
the description of these systems, the mapping procedure turns into an estimation of a few 
characteristic parameters. An asset of this technique is that it is simple to apply and 
captures the behavior of block copolymer phase separation. In our study this mapping technique 
is utilized in conjunction with a Monte Carlo (MC) algorithm to perform simulations on block
copolymer systems. The microphase separation is investigated in the bulk and under confinement, 
on unpatterned and patterned surfaces. 
\end{abstract}


\maketitle

\section{Introduction}
The combination of two or more different types of polymers as blends or copolymers
to obtain materials with improved properties is a common procedure. In most cases, however, blending
two different kinds of polymers can result in phase separation. In the case of
diblock copolymers (molecules that consist of two distinct polymer chains covalently
bonded at one end), because of connectivity constraints, the self-assembly results
into domains that exhibit ordered morphologies of a variety of ordered structures.
Depending on their chain asymmetry and the Flory-Huggins parameter, which
is a measure of the incompatibility of the different type of monomers~\cite{rubinstein},
these structures can be lamellae, cylinders, spheres, gyroids. When the self-assembly
is performed on patterned surfaces, the pattern can be used to guide the block copolymer
morphology giving precise control of the process resulting in perfect, defect free
structures that are similar to the bulk morphology or completely new structures.
These morphologies are especially suited for a number of applications in nanofabrication
such as nanowires~\cite{lopes}, photonic crystals~\cite{edrington}, quantum dots~\cite{park},
magnetic storage~\cite{cheng}.

Several theories have been proposed to study these materials. The classical theoretical
method to treat these systems is based on the Flory and Huggins lattice mean-field
model~\cite{rubinstein}. This theory has been the basis of considerable efforts to develop
methods able to describe the phase separation of polymer blends. One such popular method
is the self consistent field (SCF) technique~\cite{fredrickson}. However, the SCF theory is
built on a number of assumptions. For example, the SCF theory employs a Gaussian model for
polymer chains to describe the entropy of the system, and it adopts a Flory-Huggins interaction
parameter to account for energetic contributions to the free energy of the system. This theory
does not take into account fluctuation effects. Recently though, a new particle based SCF
method has been developed~\cite{muller2,daoulas,daoulas3} which retains the advantages of the
SCF but it also includes fluctuation effects and maintains explicitly information about the
molecular conformations.

Molecular simulations such as Monte Carlo (MC) and molecular dynamics (MD) have
been utilized in the past for the investigation of block copolymer self assembly.
Computer simulations play a significant role in testing theoretical models and in
interpreting experimental results. A study of these systems on a molecular level
is necessary and simulations are an indispensable tool to visualize the molecular 
mechanism of the self-assembly of the block copolymers. They are able to explain 
and predict the three dimensional final geometry and give details on chain conformational 
properties. These techniques highlight the role of fluctuations and provide insights 
into the local structure. Another important feature of molecular simulations is that 
structural and thermodynamic quantities are simultaneously accessible and parameters 
can be varied independently. They also provide the exact solution of a model which can 
be fully atomistic or coarse grained. A fully atomistic description of block copolymer
chains is yet computationally prohibitive, if we desire to study chains of sufficiently
large molecular weight on length scales which usually range from the nanometer to
micrometer regime and enormous time scales required for equilibration. But one more
important issue is the fact that the results are quite sensitive to the intermolecular
potentials which is a serious limitation~\cite{heine}.

A coarse grained approach is ideal due to the fact that only important details
are kept and others, such as chemical details, are neglected. Of course the
issues that exist with the choice of the potential for the atomistic description
are transferred to the coarse grained model if results obtained from atomistic
simulations are used for the mapping procedure. To circumvent this problem we
can use as an input thermodynamic or structural data from available
experimental results of the system that we choose to study rather than atomistic
simulations. An important issue that has to be noted though is the fact that
results from the coarse grained model can be predictive only to the extent of
the model parametrization.

A common technique employed when using coarse grained models is the lattice Monte
Carlo~\cite{kikuchi,larson,wang1,wang2,chen,martinez}. This algorithm has the
advantage to be computationally convenient and fast and give the ability to simulate
large systems. However, these models can introduce artificial spatial anisotropies.
An additional compromise is that the systems must be studied in an isochoric ensemble
while some problems are more easily investigated in an isobaric ensemble.

An extension of the simple lattice Monte Carlo simulations is the bond
fluctuation model. The chains are more ``flexible'' since the monomers
are not restricted only to the closest lattice sites. This technique
has been vastly used and significant calculations on phase separation
of blends and block copolymers both in the bulk and under confinement
have been performed~\cite{geisinger,muller,reister}.

Apart from Monte Carlo, molecular dynamics have also been utilized
to study block copolymer self-assembly. One method used to
investigate the phase behavior of diblock copolymers is the
discontinuous molecular dynamics~\cite{schultz,alsunaidi,groot}.
This technique converges quite fast but at the same time is limited to
small chain lengths due to equilibration efficiency. One more
strength of the discontinuous molecular dynamics is its capability
to predict the dynamical pathway along which a block copolymer
melt finds its equilibrium structure after a temperature quench.
Going further in simulation complexity, Grest et al.~\cite{grest,murat}
have performed molecular dynamics simulations aided by Monte Carlo
identity exchange move focusing on the study of block copolymer chains
in the disordered and lamellae phase in the bulk.

In this manuscript we present a facile and incomplex procedure to map
the parameters of a coarse grained model to theory and connect our
results to specific experimental systems. After obtaining the values of
the necessary parameters we utilize a Monte Carlo algorithm in continuum
with specific moves that are proven to be efficient for the study of
these systems~\cite{depablo,yioryos}. We turn to continuum simulations
to avoid problems that cannot be dealt with lattice Monte Carlo. A few
of the advantages of resorting to continuum description are the ability
to use an isobaric ensemble, the investigation of systems such as branched
or crosslinked polymers since an off lattice simulation would be more
realistic and also the study of the effect of volume differences between
species. In a continuum description we have better means to investigate
local properties and interphase local structure. Upon describing the
suggested mapping methodology, our bead spring model will be used to
reproduce experimentally retrieved results such as lamellae and cylinder
formation on unpatterned and patterned surfaces.

\section{Simulation Details}
The segments of the polymer molecules interact pairwise via the
12-6 Lennard-Jones truncated potential energy function, shifted at
the cutoff $r_{c}=2^{1/6}\sigma$. This renders the segments repulsive
to one another and the cutoff is small so that the number of
interacting neighbors is quite small.

\begin{equation}
U_{nb}(r)= \left\{ \begin{array}{ll} 4\varepsilon\left[(\frac{\sigma}{r})^{12}-(\frac{\sigma}{r})^{6} \right] - U_{LJ}(r_{c}),  & r\leq r_{c} \\
 \\
 0,   &r>r_{c} \end{array} \right.
\end{equation}
where $\varepsilon$ and $\sigma$ are the Lennard-Jones parameters
for the energy and the length respectively and $r$ the distance
between the segments. The parameter $\sigma$ has a value of $\sigma = 1$
while $\varepsilon$ is an adjustable parameter. The bonding energy
between consecutive segments in the same chain is given by
\begin{equation}
U_{b}(r)=\frac{1}{2}k\left( r-\sigma\right)^{2}
\end{equation}
with bond constant $k = 2\cdot10^{3}\varepsilon/\sigma^2$.

The surface potential is described by
\begin{equation}
U_{surf}(r)= \pm \frac{\Lambda f(x,y)}{\epsilon R_{e}} exp( \frac{-z^{2}}{2(\epsilon R_{e})^{2}})
\end{equation}
where plus and minus signs correspond to PS and PMMA substrate
interactions. The coefficient $\Lambda$ characterizes the strength
of the interaction while the function $f(x,y)$ provides the pattern
of the substrate.

The systems studied consist of chains of $N=32$ beads and the density
was chosen to be $\rho=0.7$ for convenience of computation. The
temperature was selected to be $T=2.0$. For brevity all quantities in
the manuscript are reported in LJ reduced units.

Specific Monte Carlo (MC) moves have been utilized to overcome severe
limitations that traditional MC and molecular dynamics methods are
facing in macromolecular systems. Random monomer displacements are
used for local movements together with reptation moves in a configuration
bias scheme to increase performance. While reptation can be effective in
dilute systems of short chains, for intermediate to long chain molecules
it is essential to resort to trial moves capable of rearranging inner
segments of the polymer. This is particularly important in our present
study, where block copolymer chains get trapped after a phase starts
to form and sampling the correct structure and arrangement of long
chain molecules can be particularly demanding. For this reason double
bridging trial moves are implemented~\cite{depablo,theodorou}. This is a
chain-connectivity-altering move which consists of a simultaneous
exchange of parts of two neighboring chains. Double bridging allows
for effective equilibration of the systems under study and sampling
via configurational bias significantly enhances acceptance of the
rebridging scheme. The acceptance ratio of the random displacement
is kept at $30\%$. For the double bridging either two or one beads
are chosen to be deleted and rebuilt to increase acceptance and
depending on the system under study an acceptance ratio of $0.5$
to $1.5\%$ is obtained in the case where one bead is chosen.

\section{Mapping Methodology}
In order to be able to compare results from a molecular simulation of a coarse grained model
with theory and experiments it is necessary for a number of parameters of the
coarse grained model to be mapped. The first one is the Flory-Huggins parameter,
$\chi N$, which is a measure of the incompatibility of the segments belonging
to different blocks and determines the parametrization of the cohesive interactions.
The second is the end-to-end distance of the block copolymer chain and provides
the mapping of the length scales. The interaction strength of the surface is
regulated by the parameter $\Lambda N$ and finally the invariant degree of
polymerization $\overline{N}$ controls the strength of fluctuations.

The degree of polymerization is defined as:
\begin{equation}
\overline{N} = (\frac{\rho R_{e}^{3}}{N})^{2}
\end{equation}
For a usual block copolymer system~\cite{daoulas,daoulas2,stoy,kim} the value
of this parameter is extremely large. Due to computation limitations, we shall
not use an exact value of an experimental system for our coarse grained model;
$\overline{N}$ can be increased by either increasing the chain length $N$ or the
density $\rho$. We are more interested to show how a realistic system can be mapped
and suffice with a lower value for $\overline{N}$ which will give rise to stronger fluctuations.
These fluctuations, even if they affect the occurring morphologies they still
allow for a clear pattern formation which as will be shown later is quantitatively 
in agreement to experimental results.

The next step is to obtain the length scale for our model, which, as was mentioned, is
$R_{e}$. The value of $R_{e}$ can be easily determined from a simulation of the blockcopolymer
after the Lennard-Jones parameters have been mapped to correspond to specific $\chi N$. For
example, for a density of $\rho=0.7$, chain length $N=32$, and chain asymmetry of $f=15/32$, the
system of $\varepsilon_{AA} = \varepsilon_{BB} = 0.022$ and $\varepsilon_{AB} = 1.0$ it is
found that $R_{e} = 7.8$, while for $\varepsilon_{AB} = 10.0$ we obtain $R_{e} = 8.1$.

The strength of the surface interaction $\Lambda N$ for our simulations is obtained by
considering a PMMA-PS block copolymer of chain length $N = 32$ and
$\chi N = 36.7$ confined between two uniform substrates, one PMMA-attractive
and the other PS-attractive substrate, which form a sandwich geometry. We calculate the 
density profile of the polymer segments perpendicular to the substrates. The surface
energy is determined via the convolution of the polymer-substrate interactions and the estimated
density profile.
\begin{equation}
\Delta E = \int dz U_{surf}(\rho_{A}-\rho_{B})
\end{equation}

\begin{figure}[htp]
\begin{center}
\includegraphics*[width=7cm]{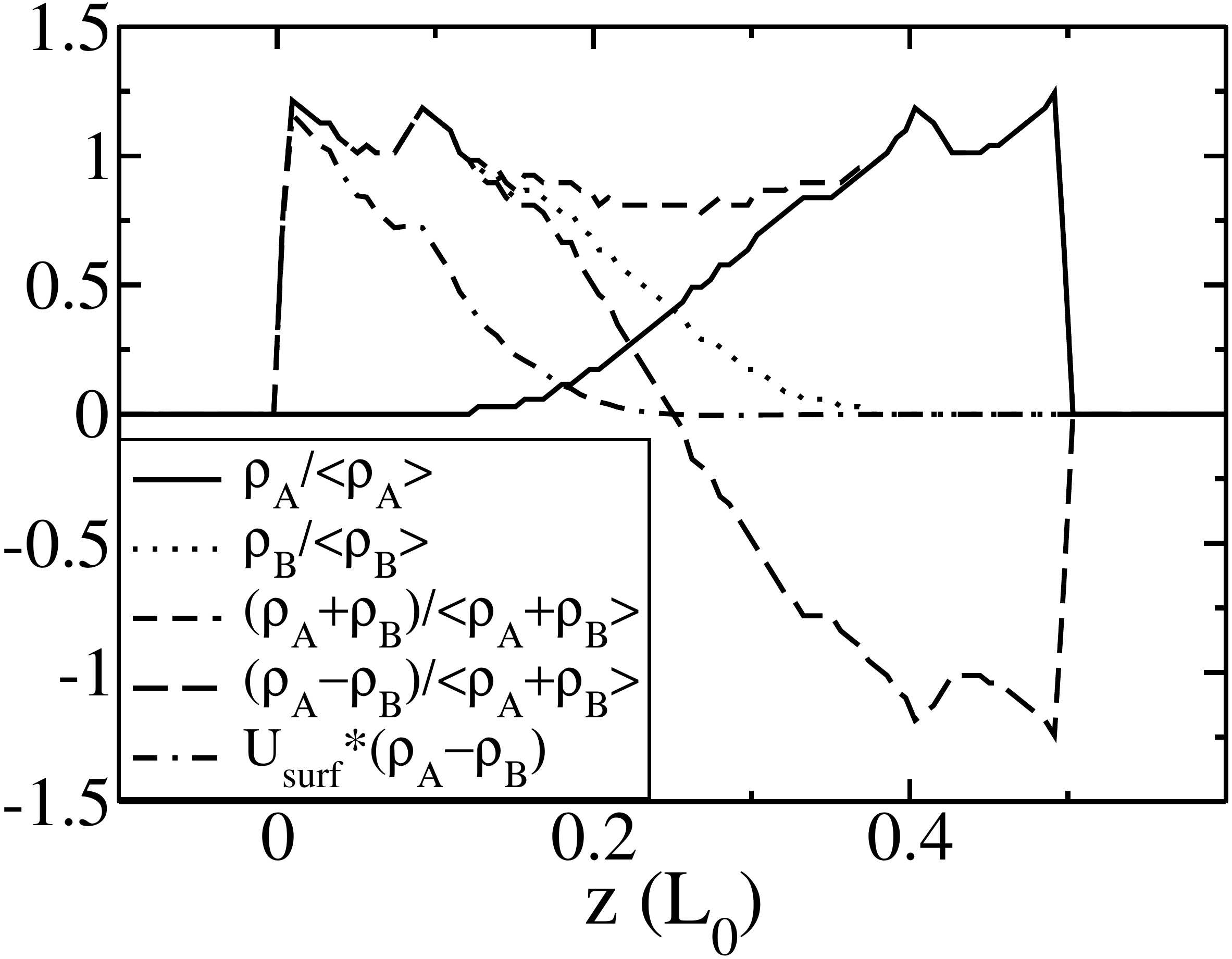}
\caption{Distribution of density for the block copolymer system in the ``sandwich'' system defined
in the text. The convolution of the density profile with the polymer-surface interaction used to
obtain the free energy differences and map $\Lambda N$ is also plotted.}
\label{figure:wallmap}
\end{center}
\end{figure}

To obtain an order of magnitude estimate for the surface free energy differences, we utilize the
adsorption energy per unit surface of a PMMA melt on a silicon oxide substrate, which as found in 
the literature~\cite{costa}, is $0.018 k_{B}T$/$nm^{2}$. In order to match this value we
choose $\Lambda N = 3.0$, which for the system $\varepsilon_{AA} = \varepsilon_{BB} = 0.022$ and
$\varepsilon_{AB} = 1.0$, corresponds to $\sim 0.016 k_{B}T$/$nm^{2}$. In Figure \ref{figure:wallmap}
the density profile and the convolution of the surface energy with the difference of density
profiles is given.

To map the Flory-Huggins parameter, $\chi N$, we perform semigrand canonical
ensemble simulations on a blend of unlike A and B homopolymers, and change the
energy interaction parameter $\varepsilon_{AB}$ of the Lennard Jones potential
between the beads of polymer chains. The chain length is chosen sufficiently
smaller in this set of simulations for efficiency. The volume fraction of the
unlike beads, from the first derivative of the Flory-Huggins free energy,
will follow:
\begin{equation}
\chi N = \frac{ln(\phi/(1-\phi))}{2\cdot\phi-1}
\end{equation}
which at the strong segregation regime reduces to:
\begin{equation}
\phi \sim exp(-\chi N)
\end{equation}

Initially chains of the same type are inserted with the aid of configuration bias
in the simulation box. When the desired density is reached, $50\%$ of the chains
are selected to be of one type and the rest of the other. The opposing type
interactions are ``switched on'' in this way. Next the system is simulated under the
semigrand canonical ensemble. By virtue of the structural symmetry of the blend components,
the semigrand ensemble reduces to an identity exchange type of move and the acceptance
criterion involves only the difference in initial and final energy. By
knowing the desired value of $\chi N$ the energy interaction parameter $\varepsilon$
can be modified until the volume fraction occurring from the simulation fulfills the
previous relation.

In the following, we will exhibit how the values of the energy interaction
parameter $\varepsilon$ of the Lennard Jones potential can be estimated so
that our bead spring model represents a system of a desired Flory-Huggins
parameter. The experimental system that will be compared is composed of
PS-b-PMMA chains of such molecular weight giving a
$(\chi N)_{exp} = 36.7$~\cite{daoulas,daoulas2,stoy,kim}. The chain length
chosen for the block copolymer self assembly simulations is $N = 32$. The
described mapping methodology is performed using smaller polymer chains
of length $4$ and changing $\varepsilon$ until the volume fraction is
\begin{equation}
\phi \sim exp(-(\chi N)_{experimental}*4/32)
\end{equation}

\begin{figure}[htp]
\begin{center}
\includegraphics*[width=7cm]{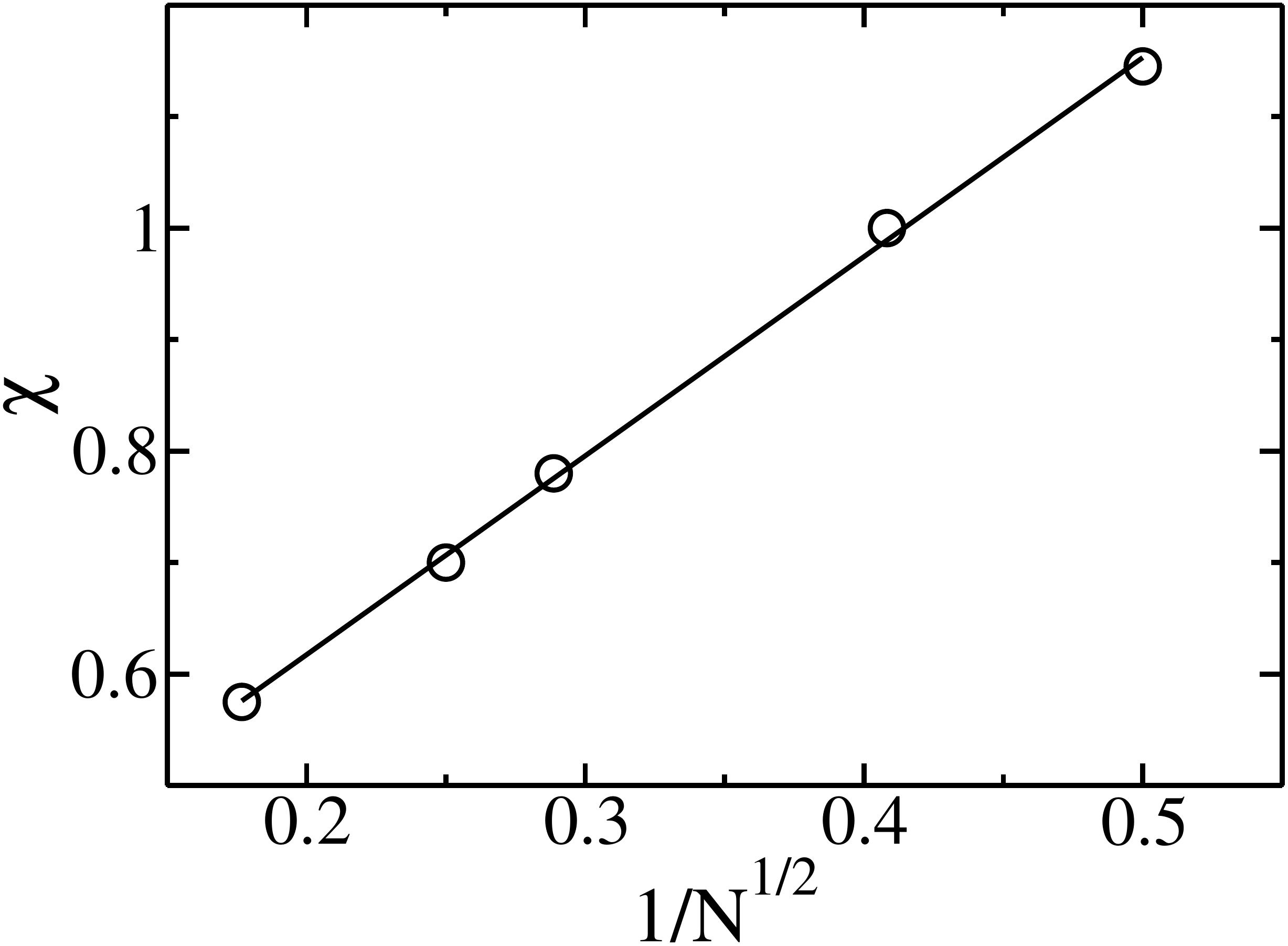}
\includegraphics*[width=7cm]{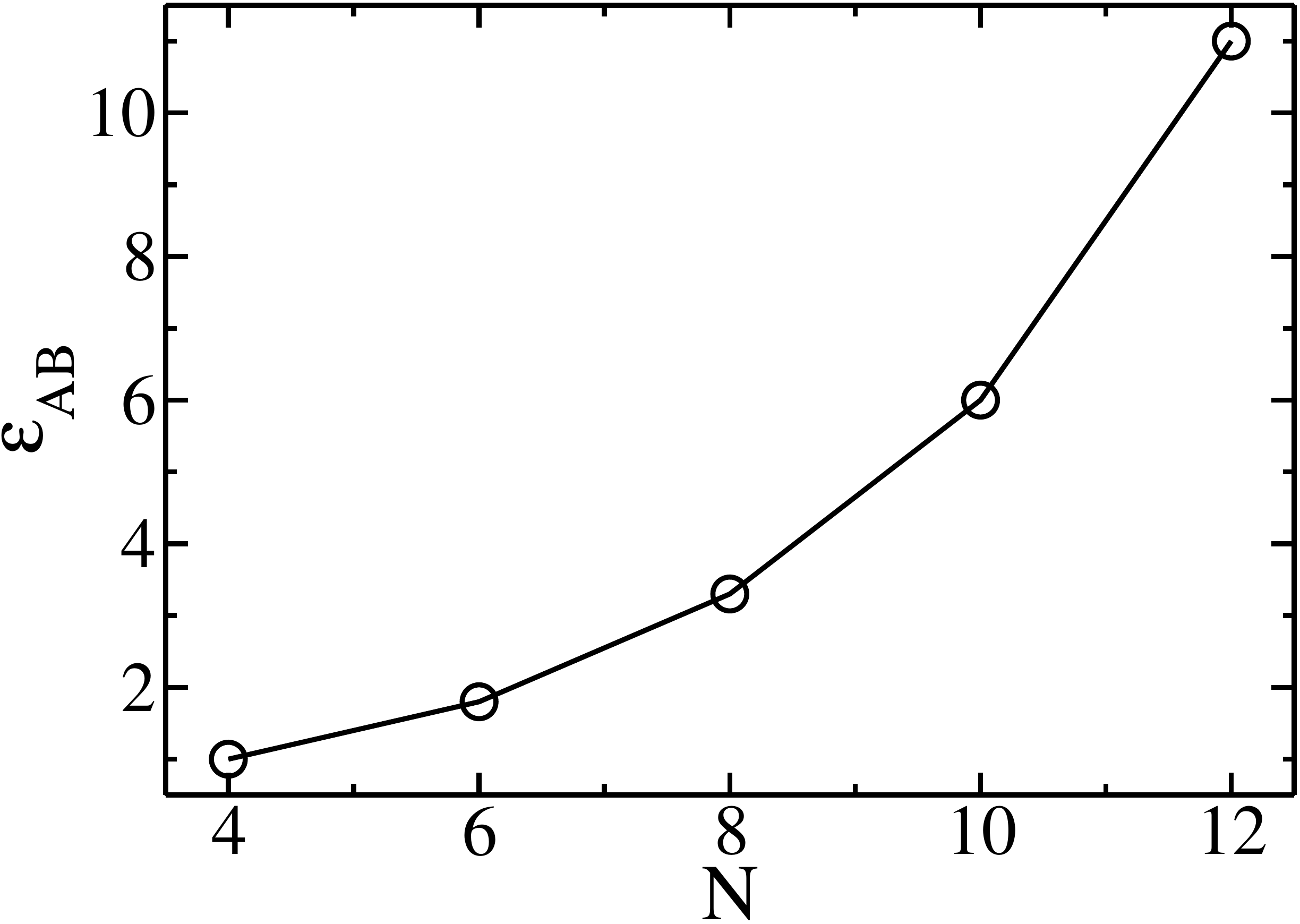}
\caption{a) $\chi$ dependence on molecular weight for a blend of unlike A and B homopolymers
system and Lennard-Jones parameter, $\varepsilon_{AB}=1$. b) Value of Lennard-Jones parameter,
$\varepsilon_{AB}$, in order to obtain $\chi N = 36.7$ for block copolymer chains of length
$N = 32$ using chain of different lengths for the mapping procedure through semigrand canonical
simulation.}
\label{figure:checkepsilon}
\end{center}
\end{figure}

From our calculations we find that the values of the Lennard-Jones energy parameters that
correspond to this $(\chi N)_{exp}$ are $\varepsilon_{AA} = \varepsilon_{BB} = 0.022$ and
$\varepsilon_{AB} = 1.0$ using the test case $N=4$. We continue by performing semigrand 
canonical simulations using the extracted parameters, but for different chain lengths in order to recover
the value of $\chi$ as a function of $N$. From Figure \ref{figure:checkepsilon}a we see that 
the value of $\chi$ depends on molecular weight. This molecular weight dependence can be 
described by~\cite{gennes,sun,kamal}:
\begin{equation}
\chi = \chi_{\infty} + k*N^{-1/2}
\end{equation}

From this result it is clear that for the calculations of the mapping procedure
we have to use polymer chains of the same length as the ones that will be used
for modeling these systems. However, these simulations are 
computationally expensive for large chain lengths as was already mentioned. We can perform instead the
calculations for a range of smaller chain lengths and then extrapolate to the desired one.
The results are plotted in Figure \ref{figure:checkepsilon}. 
We see that if this molecular weight dependence was not taken into account, then the mapping procedure would have
resulted in a set of parameters giving an incorrect value of $\chi$.
If we extrapolate the data to $N = 32$, we find that $\varepsilon_{AB} = 284$ for $(\chi N)=36.7$.

\begin{figure}[htp]
\begin{center}
\includegraphics*[width=6cm]{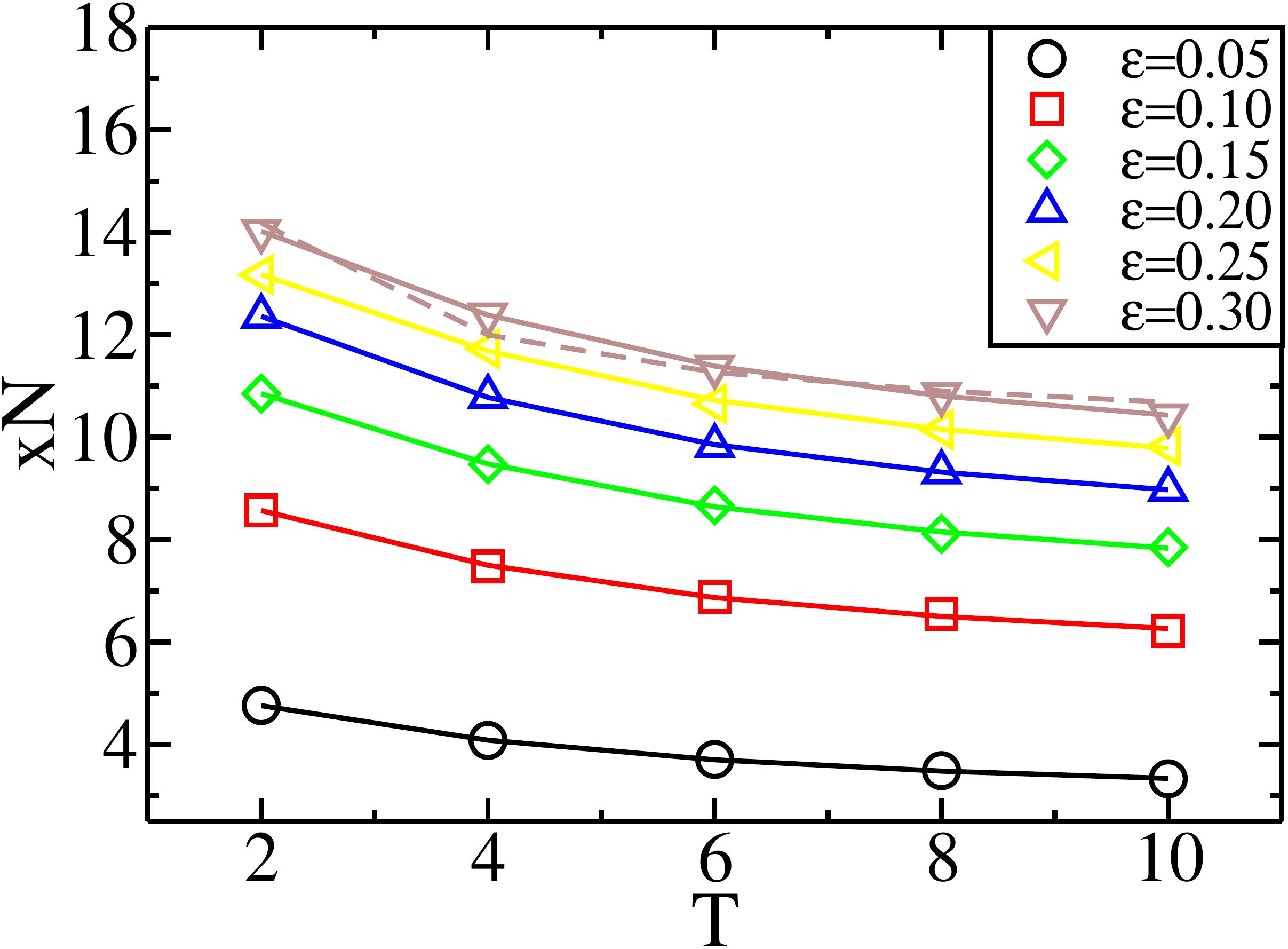}
\includegraphics*[width=6cm]{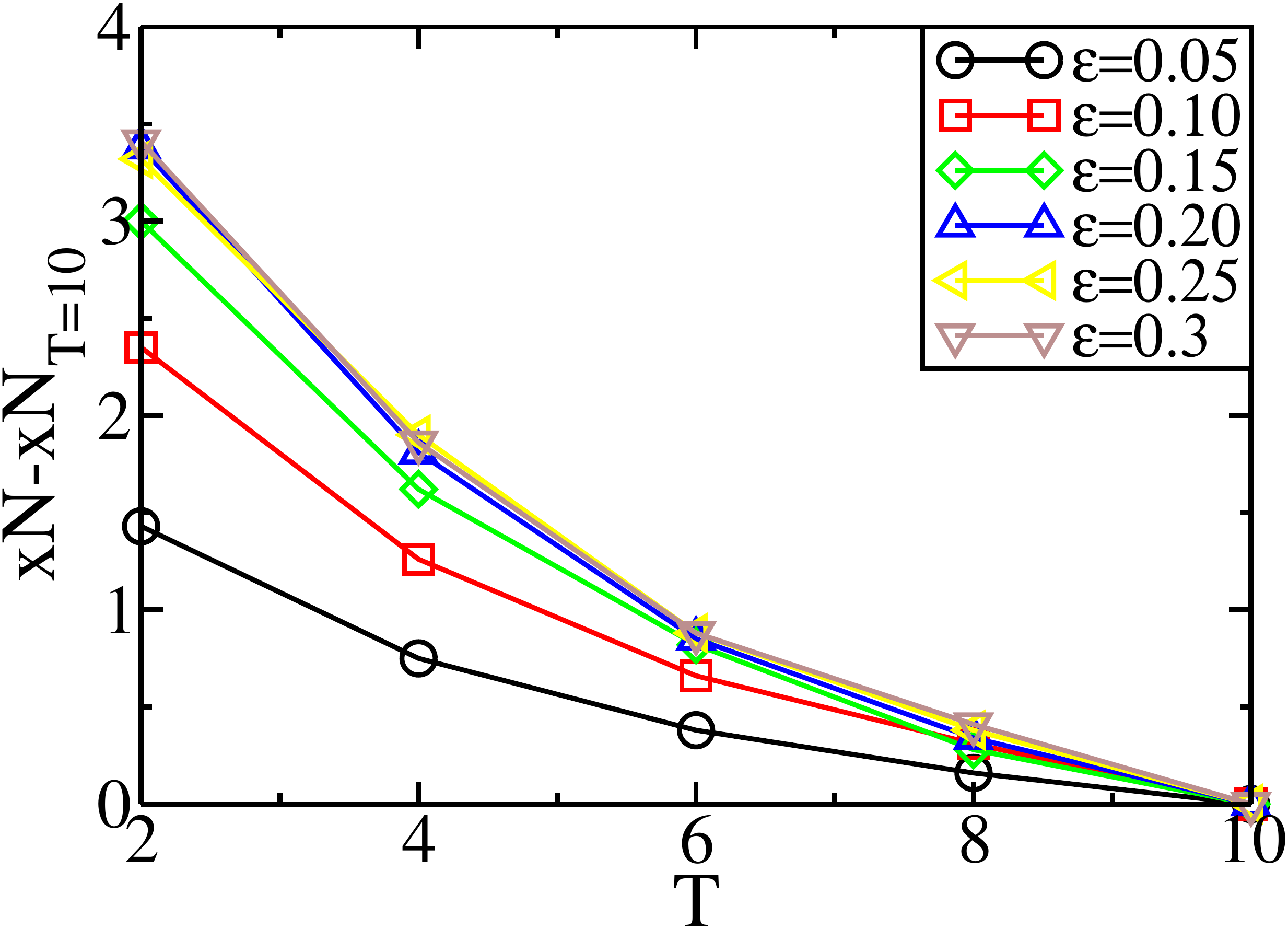}
\caption{a) $\chi N$ with respect to temperature for different values of $\varepsilon_{AB}$
and for chain length $N = 32$. For low values of $\varepsilon_{AB}$ the curve appears to be
linear. The solid lines represent fits with equation 10. The dashed line is plotted for comparison
with a linear fit of our data. b) $\chi N$ with respect to temperature for different values of $\varepsilon_{AB}$
and for chain length $N = 32$. Curves shifted in order for $\chi N = 0$ zero at $T = 10$. 
Collapse of curves of $\chi N$ with respect to temperature for high values of $\varepsilon_{AB}$.}
\label{figure:32_xN-T}
\end{center}
\end{figure}

We now examine the temperature effect on $\chi N$.
For a set of temperatures and a series of $\varepsilon_{AB}$ the value of $\chi N$ is calculated.
Our results are exactly described by a quadratic function in $1/T$ of the form~\cite{balsara}:
\begin{equation}
\chi(T) = A + B/T + C/T^{2}
\end{equation}
We thus find a non-linearity of $\chi$ as a function of $1/T$. Another interesting point is the fact
that for high values of $\varepsilon_{AB}$ the curves collapse on each other, as can be seen from 
Figure \ref{figure:32_xN-T}b. This suggests that another way for mapping $\varepsilon_{AB}$
for a system of $N = 32$ is to perform the calculations for the mapping procedure at a high $T$ and then
use the collapsed curve to extrapolate the desired value of $\chi N$ for a specific temperature.

\begin{figure}[htp]
\begin{center}
\includegraphics*[width=6cm]{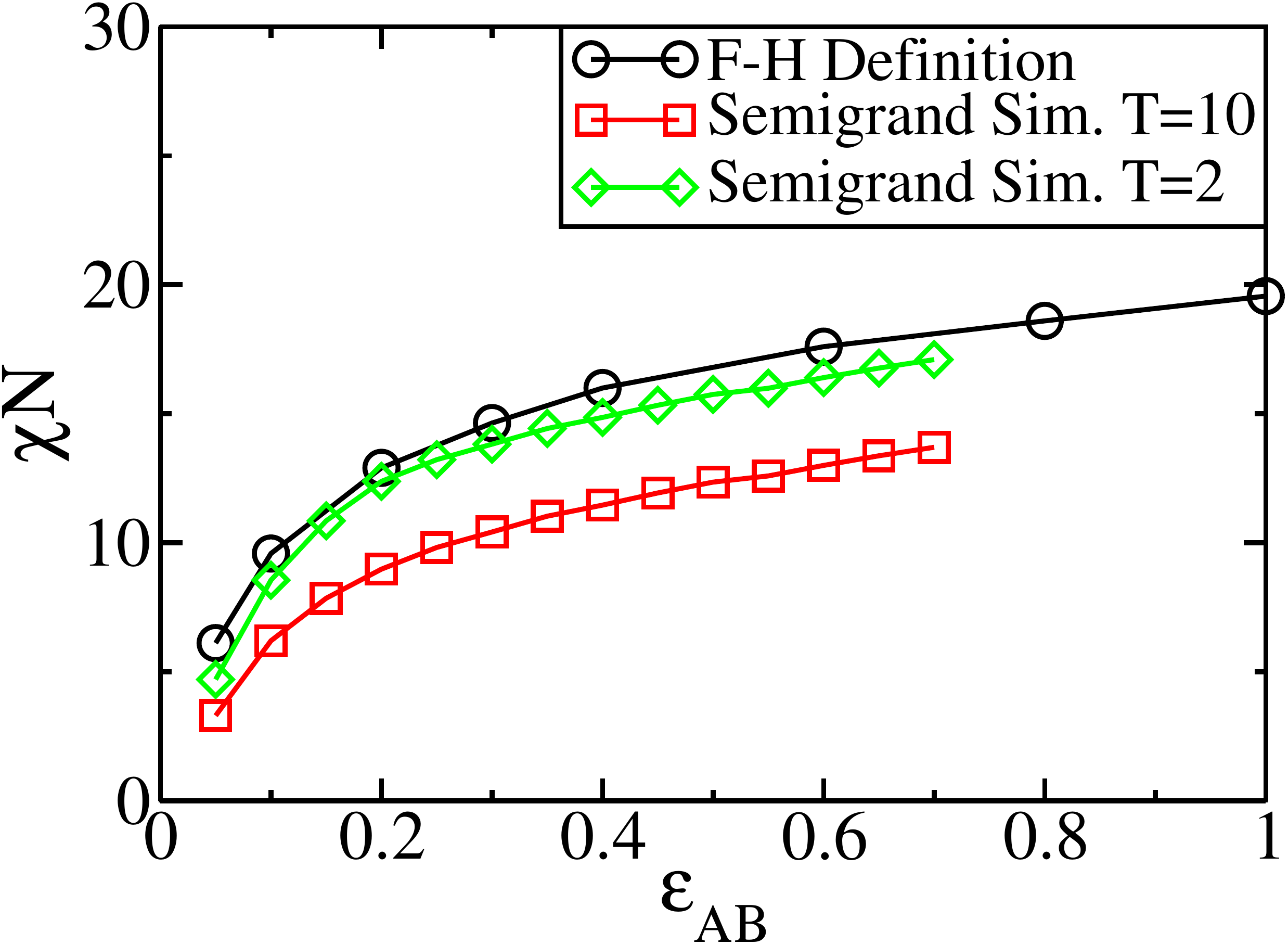}
\caption{$\chi N$ with respect to $\varepsilon_{AB}$ for chains of length $N = 32$
and for the two methodologies. The technique utilizing the semigrand canonical simulations
and the one extracting $\chi$ as defined from Flory-Huggins theory.}
\label{figure:32_eps}
\end{center}
\end{figure}

We continue by performing simulations in the semigrand canonical ensemble for a 
system of chain length $N = 32$ at temperature $T = 10$ and for a range of $\varepsilon_{AB}$.
As can be seen from Figure \ref{figure:32_eps}, $\chi N$ increases as $\varepsilon_{AB}$
increases. Using the correction from Figure \ref{figure:32_xN-T}b we obtain the corresponding
curve for $T = 2$. In addition to this mapping technique, we perform an NVT simulation of
a symmetric blend of homopolymers and calculate the value of $\chi N$ from the definition
of Flory-Huggins theory for the same range of values of $\varepsilon_{AB}$:
\begin{equation}
\chi = \frac{z*\Delta\omega}{kT}
\end{equation}
where $z$ is the coordination number (the number of nearest neighbors or the average number
of sites surrounding an individual segment below the cutoff radius), and
$\Delta\omega = \Delta U_{AB}=\frac{\epsilon_{AB}-(\epsilon_{AA}+\epsilon_{BB})/2}{\epsilon_{AB}}$
is the energy increment per A-B monomers contact (excess potential energy attributed to the A-B
interactions over the number of A-B interactions). From Figure \ref{figure:32_eps} we find
that for all the range of $\varepsilon_{AB}$ values, the difference of the two methodologies is
small. In order to verify the agreement of these two methods we perform a calculation
with the last mapping methodology using $\varepsilon_{AB}=284$ and we find $\chi N = 34.8$
in close agreement with our desired result.

We continue our study by investigating how the value of $\chi$ is affected by the composition and
the difference between having a blend or block copolymer. We use the previously described methodology, where $\chi$
is directly obtained by the definition from the Flory-Huggins theory. First we perform a calculation
for a blend system with a composition of $50\%$ and then a symmetric block copolymer system. 
Two values of $\epsilon_{AB}=1$ and $300$ are chosen.
We find for both systems that the value of $\chi$ is the same independent of system type (blend or block copolymer),
$\chi_{\epsilon_{AB}=1}\cong0.619$ and $\chi_{\epsilon_{AB}=300}\cong1.096$. This suggests that the fact that
the system is a blend or a block copolymer does not have any effect on the value of $\chi$. We perform
a calculation for blend systems of compositions $1\%$, $25\%$, and $50\%$. It is found
that again the value of $\chi$ remains constant. This is a special case for our system since the
interaction between similar type segments are considered to be identical. In general~\cite{stryuk,lee},
a composition dependence exists for the Flory-Huggins interaction parameter $\chi$.

\section{Testing the model}
As a test system we choose to study the behavior of lamellae forming A-B block copolymer chains on
patterned surfaces having the stripe geometry. The mismatch of the pattern spacing $L_{s}$ to the block 
copolymer lamellae period can affect the final structure. In order to proceed with the investigation
of this problem, we have to calculate one extra parameter: the block copolymer lamellae period $L_{0}$. 
Block copolymers chains are inserted between two hard wall surfaces using the configuration bias method. 
The Lennard-Jones interaction parameters are assigned the values 
$\varepsilon_{AA} = \varepsilon_{BB} = 0.022$ and $\varepsilon_{AB} = 1.0$ and the system is equilibrated
(the extraction of these parameters has been shown in the previous Chapter). 
The block copolymer chain length is $N=32$ and comprised of $15$ beads of A and $17$ beads of B copolymer
in order to match chain asymmetry of experimental results~\cite{kim} for PMMA-PS blockcopolymer.
An initial guess for $L_{0}$ comes from the literature~\cite{kim,sferrazza} and is $1.7$ $R_{e}$. 

The z axis of the simulation box is chosen to be normal to the two surfaces. We perform a $NPT$
simulation setting the x direction to be multiple of $L_{0,guess}$. The pressure is chosen
so that the average density of the system is $\rho = 0.7$. Only the x and y box lengths are allowed
to change keeping the z length constant. The system is equilibrated and the lamellae period is 
found to be $L_{0}=14.1\sigma$ or $L_{0} = 1.81 R_{e}$.
\begin{figure}[htp]
\begin{center}
\includegraphics*[width=7cm,angle=0]{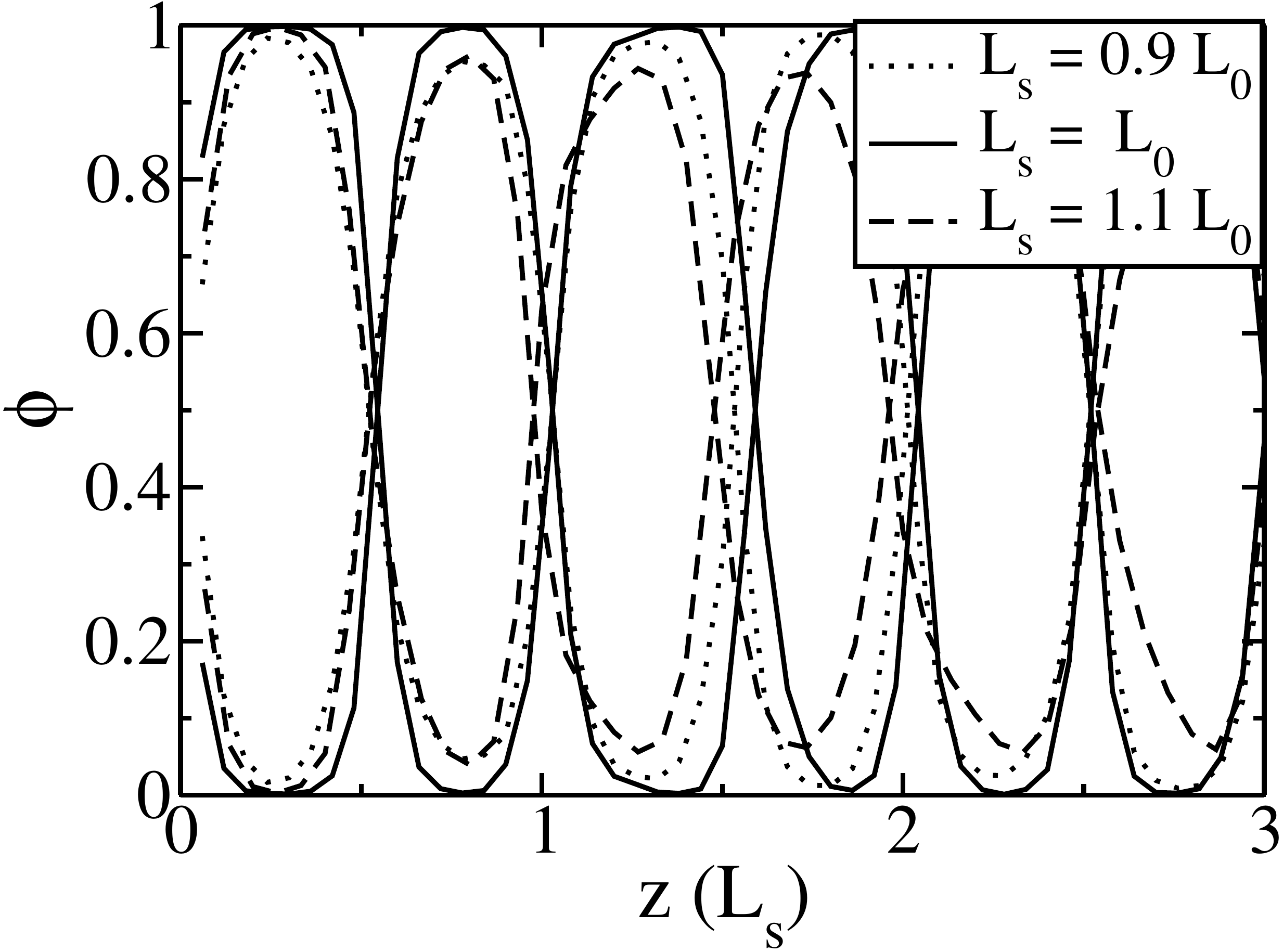}\\
\includegraphics*[width=4cm,angle=90]{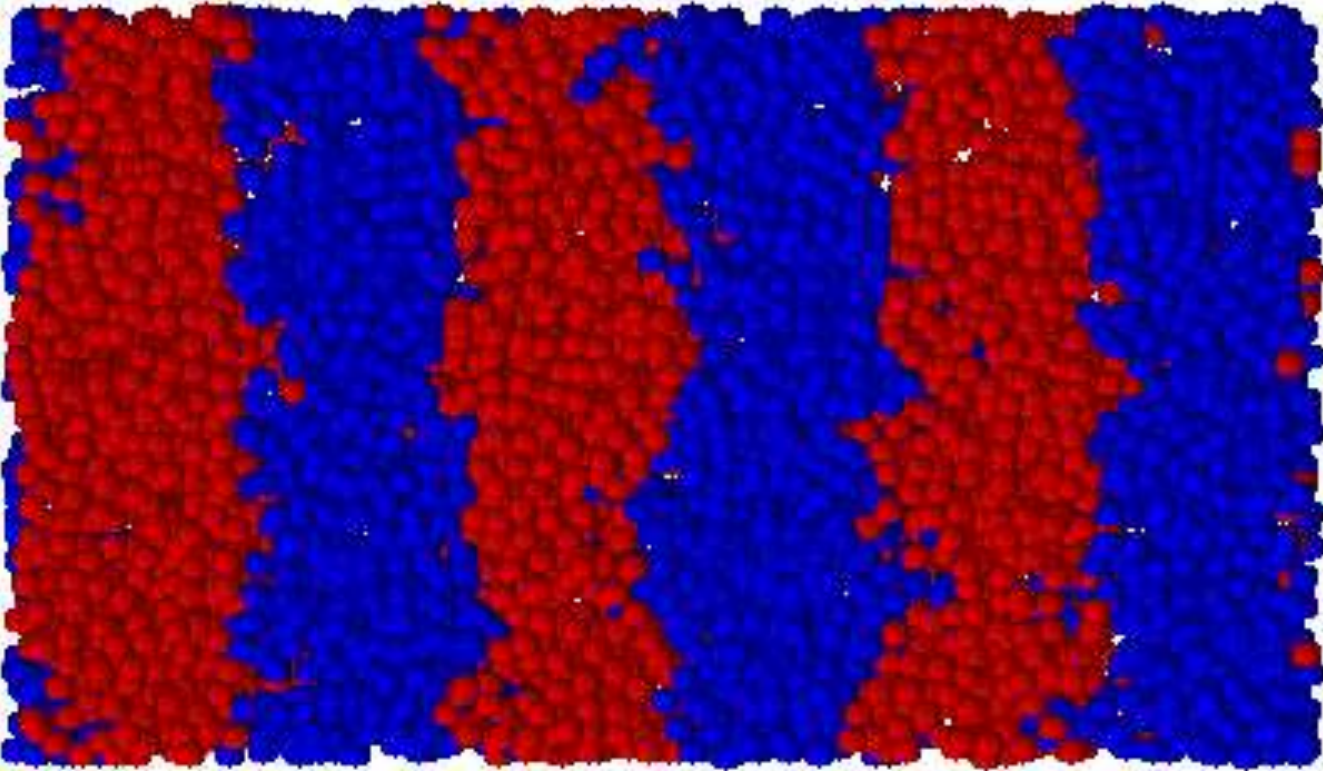}\\
\includegraphics*[width=2cm,angle=270]{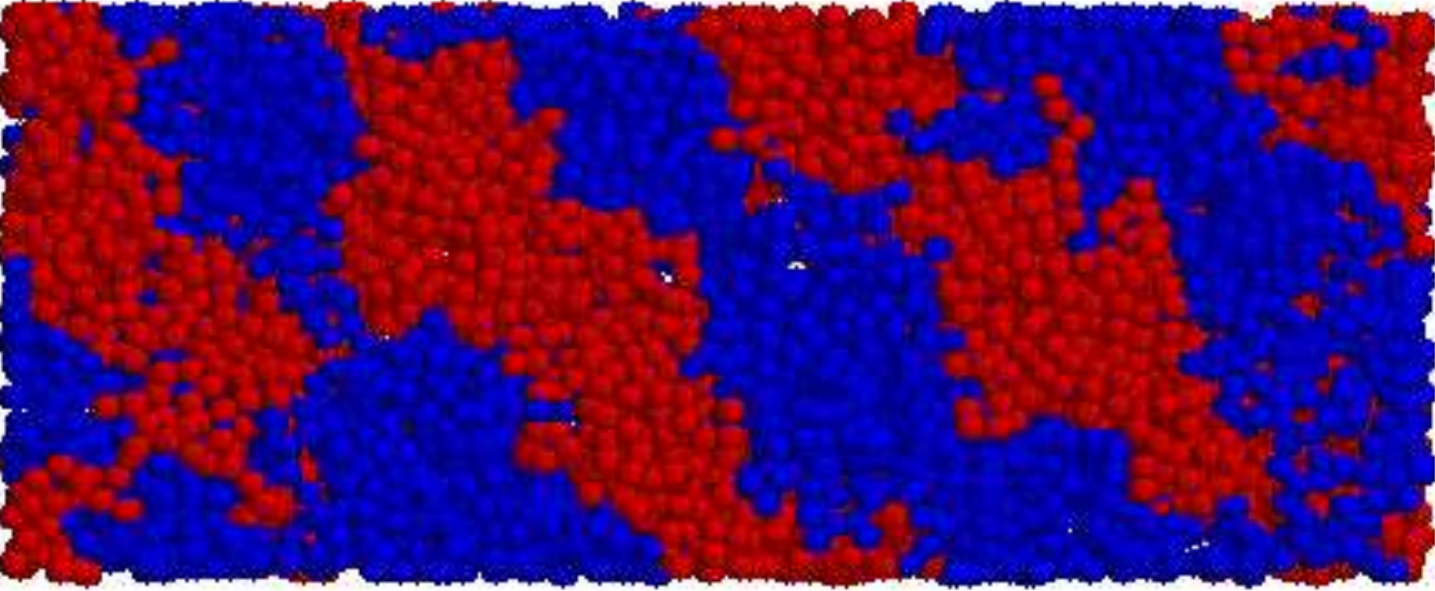}
\includegraphics*[width=2cm,angle=270]{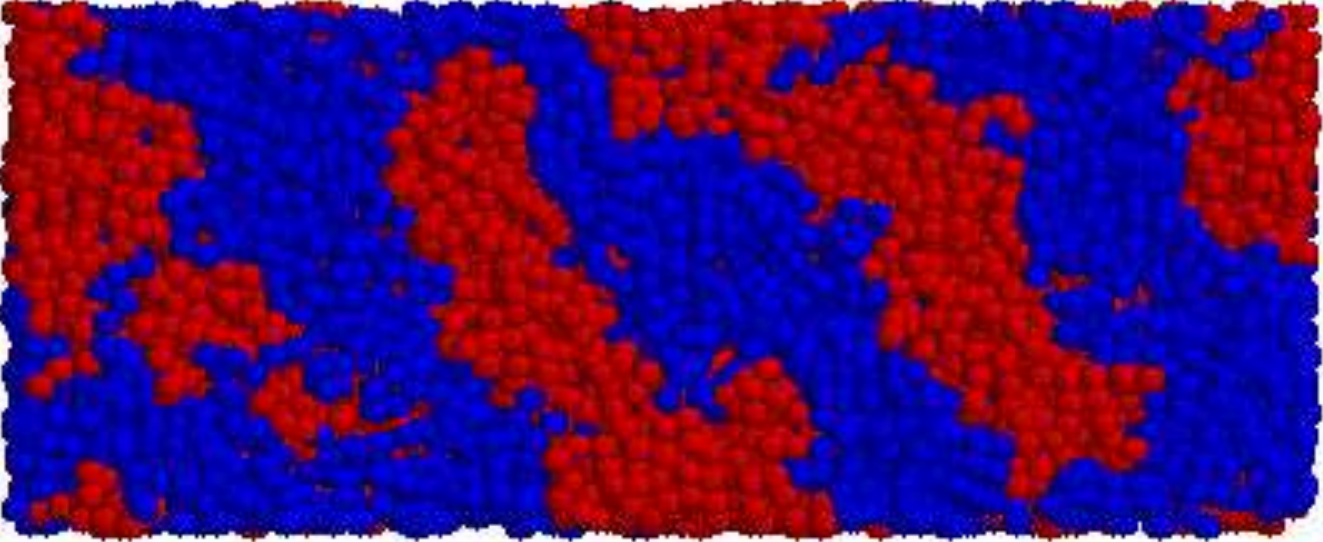}
\includegraphics*[width=2cm,angle=270]{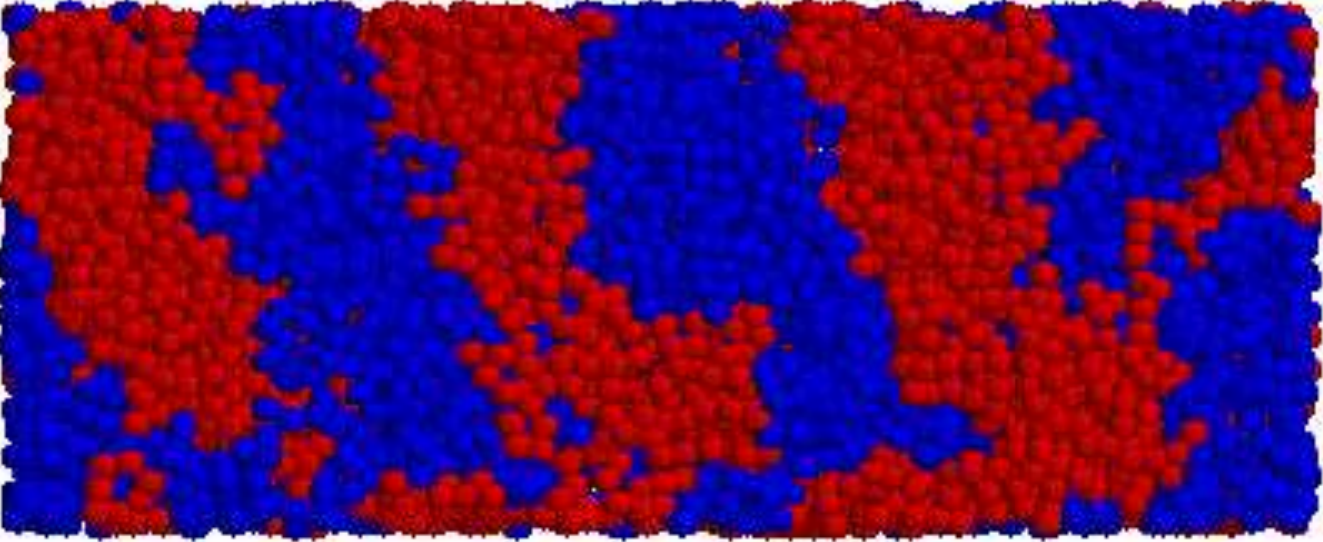}
\caption{a) Distribution of volume fraction of a block copolymer on a stripe patterned surface. Results
for mismatch of the pattern spacing and the block copolymer lamellae period are plotted for both extensions
and compressions. b) Top view snapshot of the block copolymer film simulation on a stripe patterned
surface with pattern spacing commensurate to the lamellae period and $10\%$ mismatch. c,d,e) Consecutive sideview snapshots 
of block copolymer with an increase of pattern spacing of $20\%$.}
\label{figure:bcbetweensurfaces}
\end{center}
\end{figure}

We can now proceed with the study of the previously stated problem. The block copolymer 
is deposited on a patterned surface of stripe geometry with pattern $L_{s}$. A mismatch 
is chosen between the pattern spacing and the blockcopolymer period ranging from $-20\%$ 
to $+20\%$. Upon equilibration of the systems the density profile along the normal 
to the lamellae axis is calculated. The results are plotted in Figure \ref{figure:bcbetweensurfaces}a.

Lamellae are obtained for the full spectrum of the previously mentioned ``strains''.
We find that for both compression and extension, a certain ``strain'' can be tolerated 
as seen in Figure \ref{figure:bcbetweensurfaces}b. Below or above these critical ``strains'', 
defects arise that are ``frozen'' as shown in Figures \ref{figure:bcbetweensurfaces}c,d,e where
subsequent sideview snapshots of the system are plotted. The simulation time between 
these last configurations is equal to five rouse relaxation times. Similar defects have been observed 
by experiments under SEM and AFM~\cite{kim} and quantitative agreement is found with our simulations.

The systems that we investigated so far consisted of block copolymer chains with a chain
asymmetry and a $\chi N$ value that corresponds to the lamellae region of the phase
diagram. The natural next step to test our model is to proceed with the investigation of
another domain of the phase diagram and more specifically the cylinder regime. We choose 
values of chain asymmetry and Lennard-Jones interaction parameters that result in the cylinder phase.
A chain length of $N = 32$ is chosen with $22$ segments of A type and the rest $10$ of B type 
giving a fraction of $0.3125$. The Lennard Jones energy parameters are chosen to be
$\varepsilon_{AA} = \varepsilon_{BB} = 0.022$ and $\varepsilon_{AB} = 19.0$, which
correspond to a $\chi N$ of $30.6$.

\begin{figure}[htp]
\begin{center}
\includegraphics*[width=6cm,angle=90]{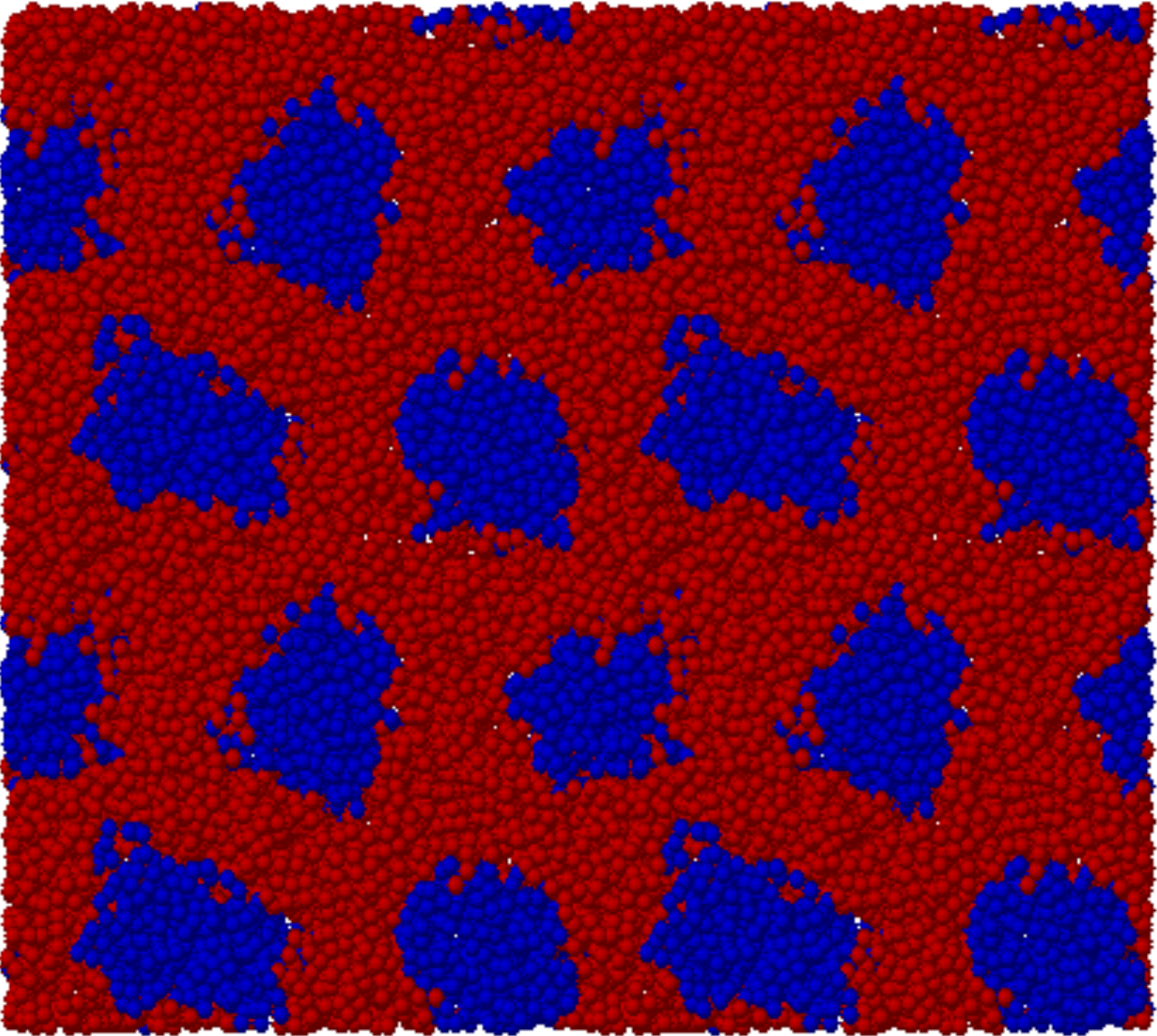}
\caption{Thin film of asymmetric block copolymer chains forming a hexagonal cylinder morphology.}
\label{figure:cylinder}
\end{center}
\end{figure}

Block copolymer chains are inserted between neutral surfaces. A simulation in the $NPT$ 
ensemble was performed changing the axes independently. After equilibration, it is found 
that $L_{x} = sin(\frac{\pi}{3})*L_{y}$. This relation is necessary in order for the system
to be able to accommodate the hexagonal structure of the cylinder phase, which can be seen from 
geometrical calculations. The cylinder to cylinder spacing was measured to be $L_{cc} = 2.1 R_{e}$. 
The resulting two dimensional density profile is plotted in Figure \ref{figure:cylinder}. The
hexagonal cylinder morphology is clearly observed.

We continue our analysis by investigating the behavior of this cylindrical 
forming block copolymer on striped nanopatterned surfaces were each
stripe is preferential to one block and repulsive to the other. The two
stripes are symmetric with $W = \frac{L_{S}^{A}}{L_{S}} = 0.5$ where
$W$ is the stripe asymmetry, $L_{S}^{A}$ the thickness of a stripe
preferential to the $A$ part of the block copolymer chain and $L_{S}$
the thickness of the pattern period. The pattern periods of the surface
were chosen equal to the cylinder to cylinder spacing. The thickness
of the film was varied and the equilibrium morphologies were obtained.
Recently, Edwards et al.~\cite{edwards2} have investigated these systems
experimentally and have characterized with scanning electron microscopy
the top part of these polymer films. However, the rest of the film cannot
be characterized with these experimental techniques. With our simulations 
we attempt to verify and complement this analysis.

\begin{figure}[htp]
\begin{center}
\includegraphics*[width=4cm,angle=270]{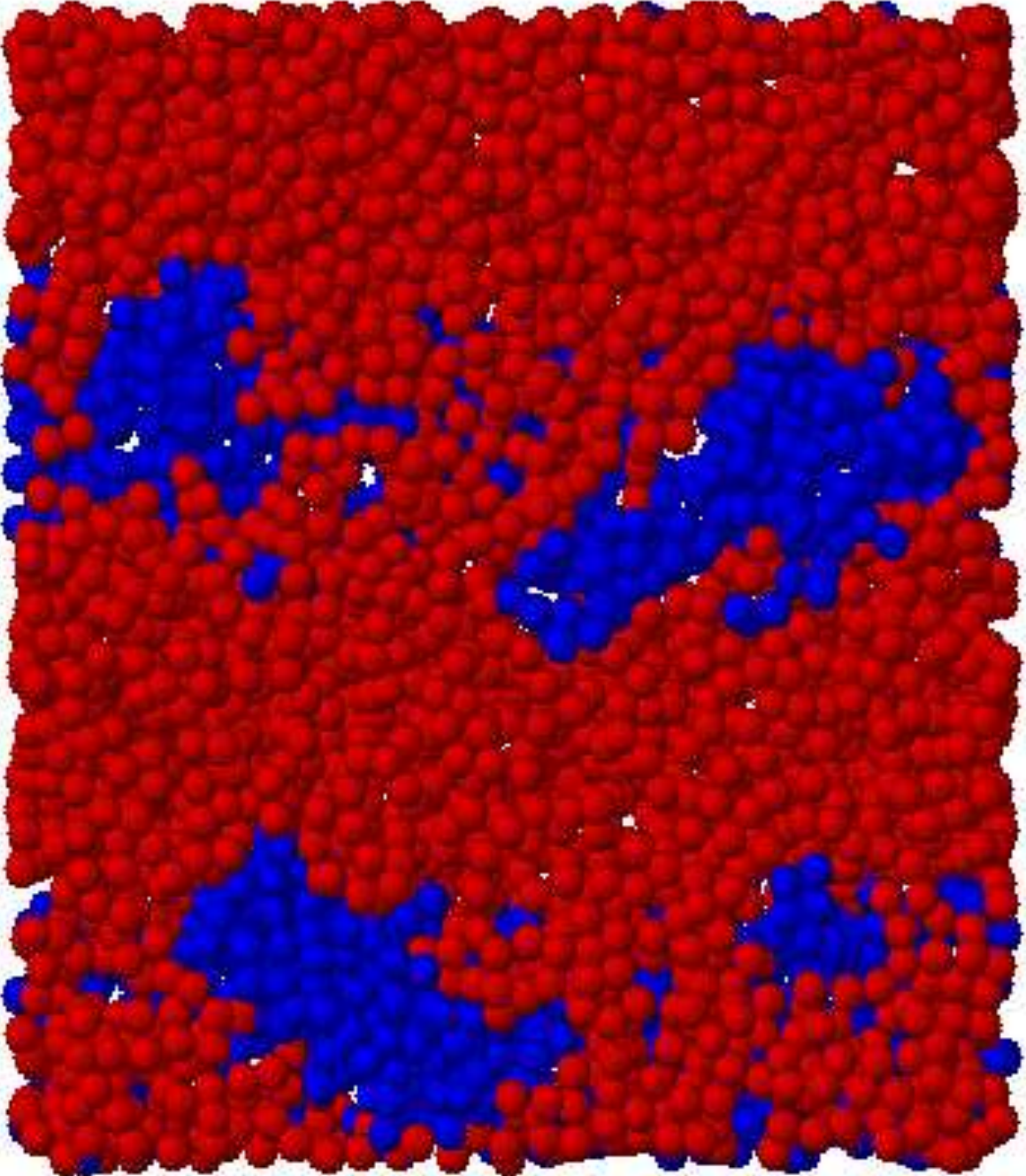}
\includegraphics*[width=2.1cm,angle=270]{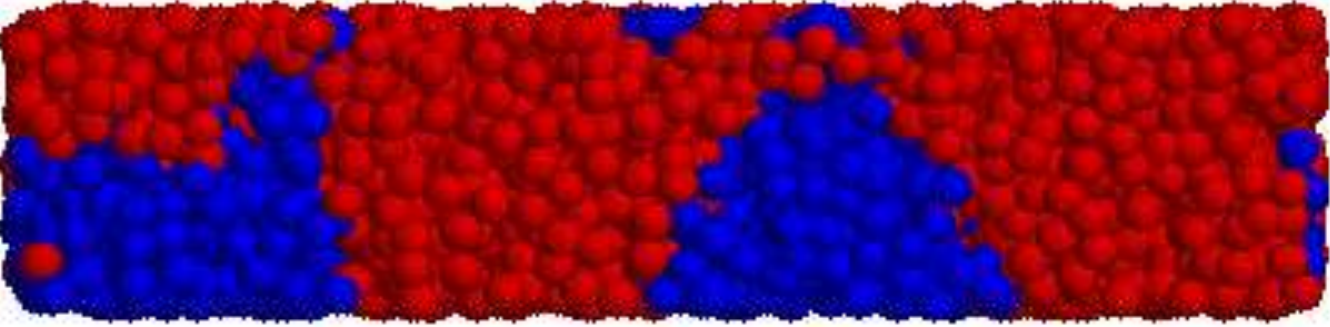}
\includegraphics*[width=4cm,angle=270]{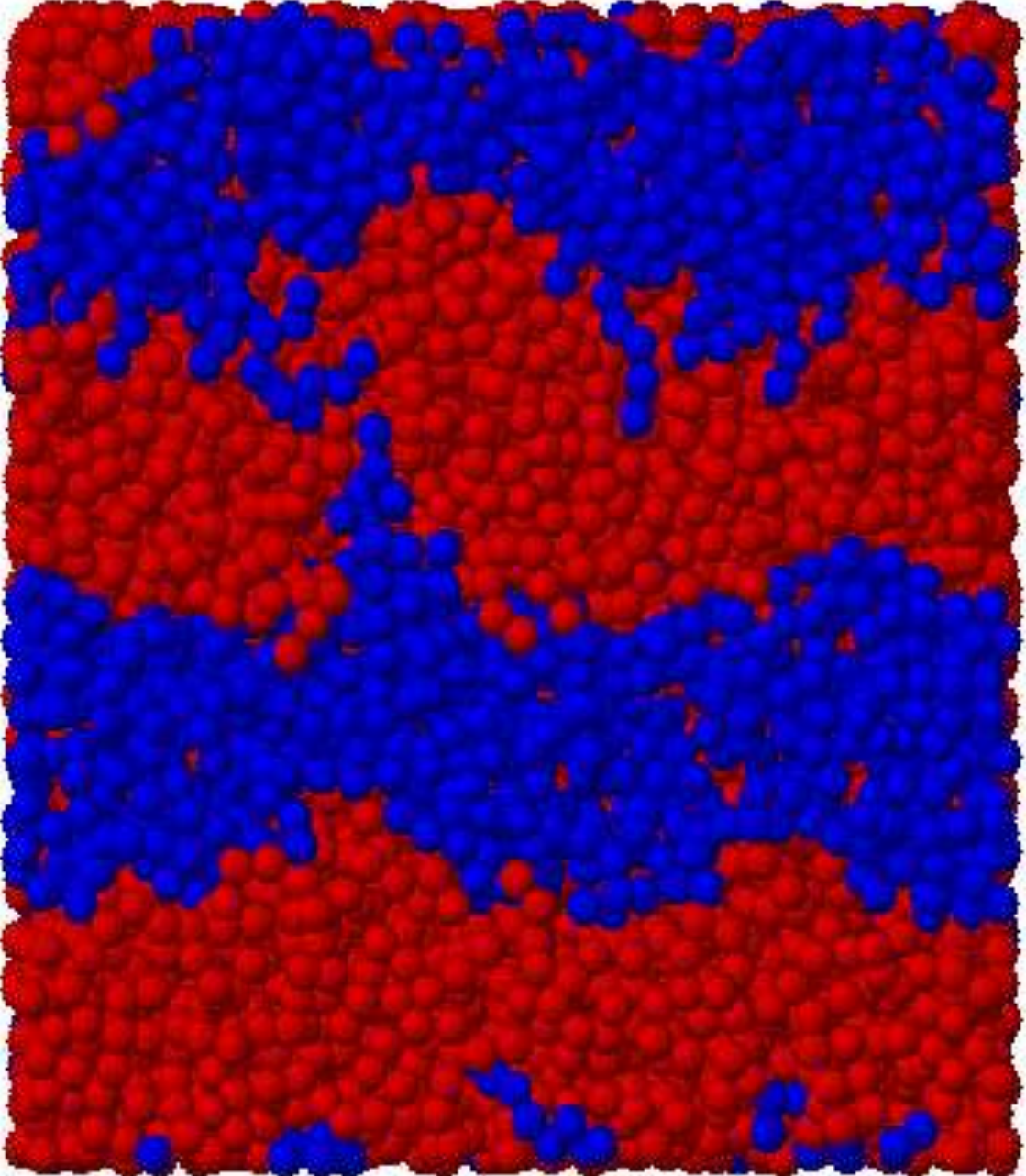}
\includegraphics*[width=4cm,angle=270]{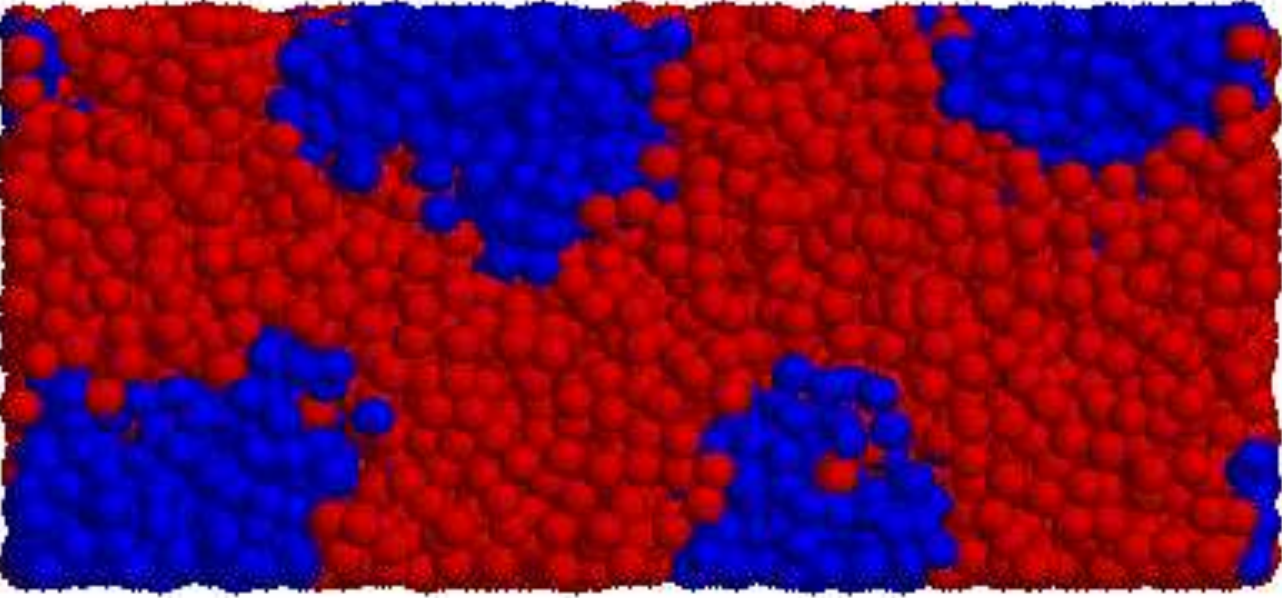}
\caption{Morphology change of asymmetric block copolymer chains on patterned
surfaces, with pattern periods equal to the cylinder to cylinder spacing and
varying the height of the film. The top and sideview of the systems is plotted.
Thickness: a,b) $7 \sigma$ or $L_{cc}\cdot sin(\frac{\pi}{3})/2$, c,d) $13.6 \sigma$ or
$L_{cc}\cdot sin(\frac{\pi}{3})$}
\label{figure:height}
\end{center}
\end{figure}

In Figure \ref{figure:height} the morphologies for different film heights are given.
As we can see for a thickness commensurate to $L_{cc}\cdot sin(\frac{\pi}{3})/2$,
semicylinders are formed on the stripe patterns (Figure \ref{figure:height}a,b).
Increasing the film thickness, making it commensurate to $L_{cc}\cdot sin(\frac{\pi}{3})$,
the block copolymer thin film assembles to form a layer of defect free semicylinders
at the patterned surface as previously and a second layer of semicylinders at the
top surface both with the same repeat period equal to the cylinder to cylinder
spacing and the chemical surface pattern (Figure \ref{figure:height}c,d).

\begin{figure}[htp]
\begin{center}
\includegraphics*[width=4cm,angle=270]{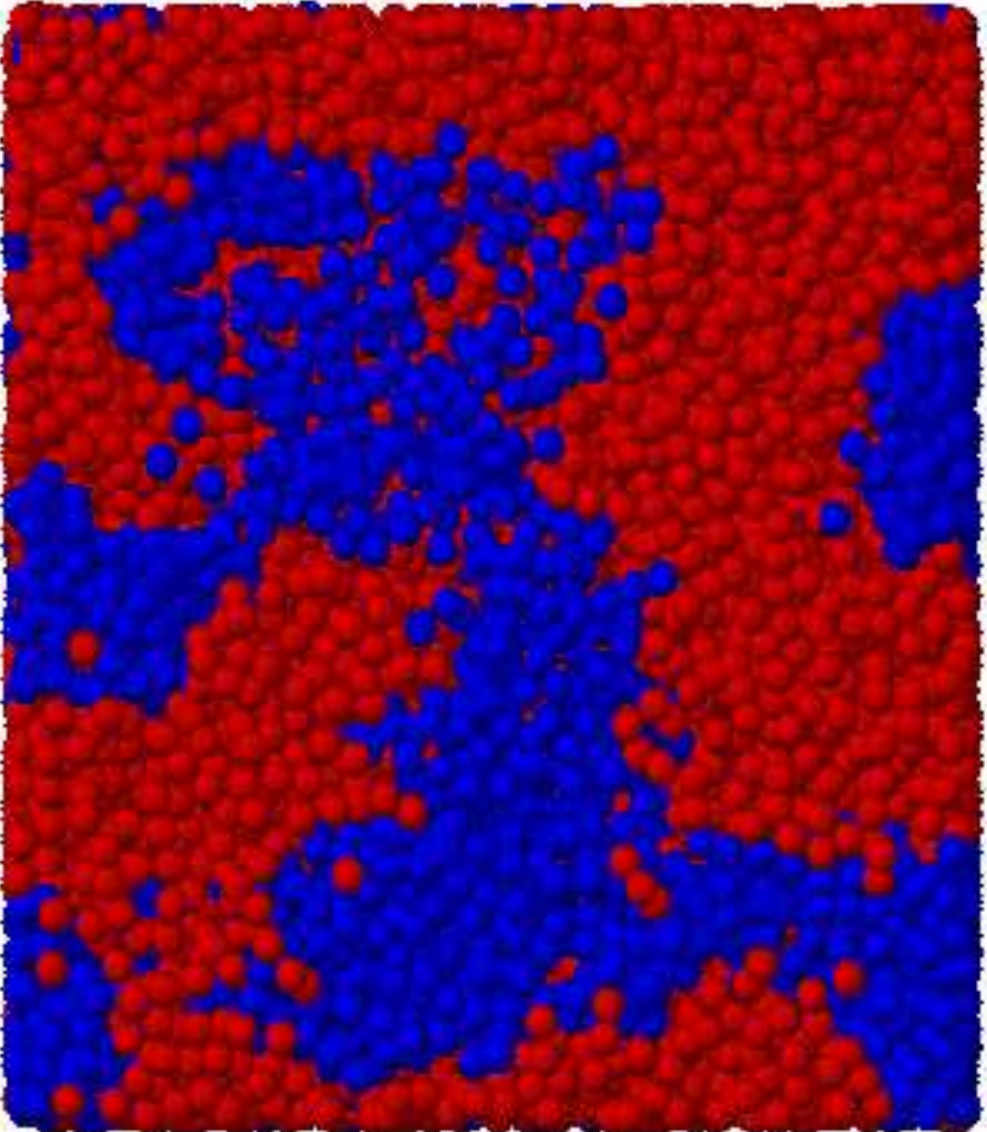}
\includegraphics*[width=4cm,angle=270]{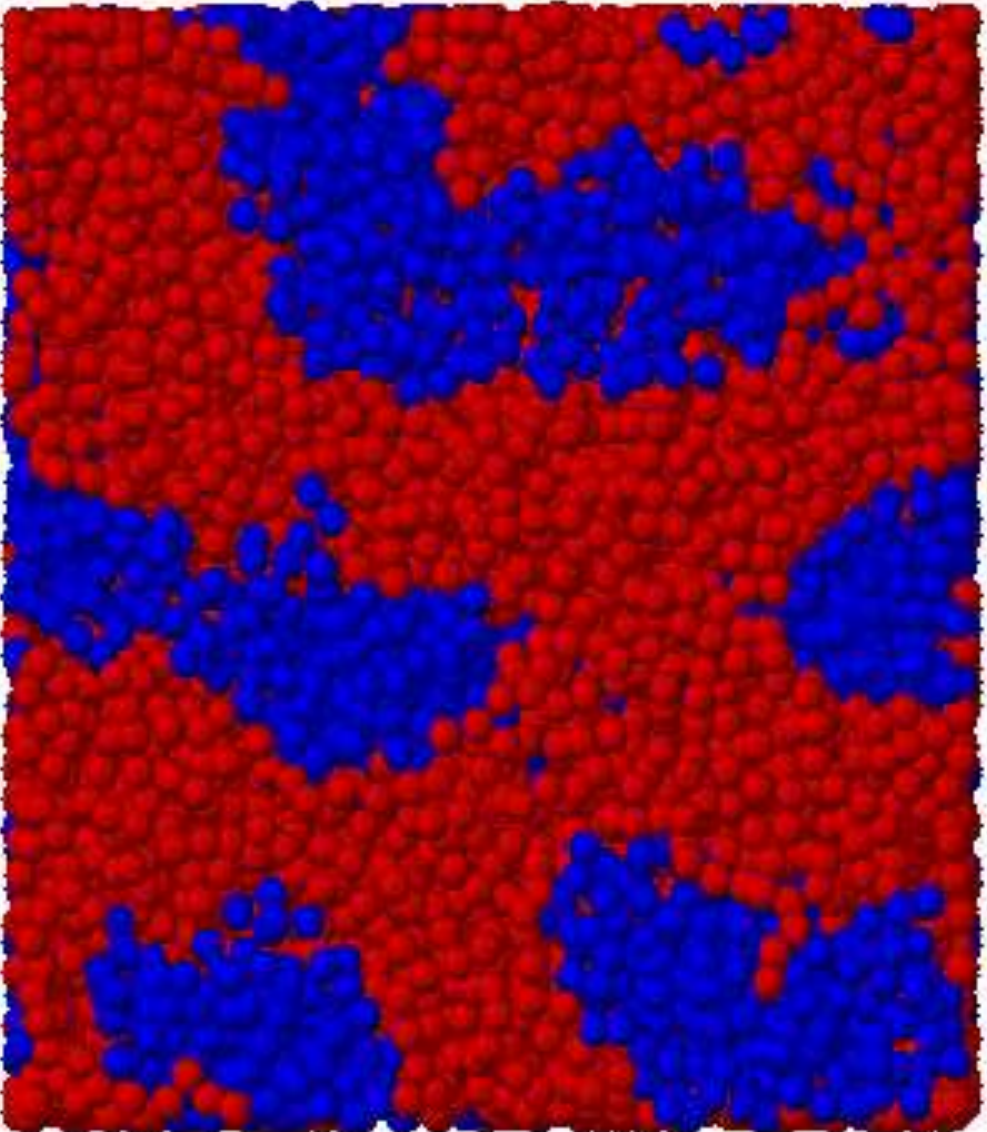}
\caption{Morphology change of asymmetric block copolymer chains on patterned surfaces, with
pattern periods equal to the cylinder to cylinder spacing for higher film thickness.
a) Sideview of a systems of thickness $27.2 \sigma$ or $2 \cdot L_{cc}\cdot sin(\frac{\pi}{3})$
b) Topview of a systems of thickness $27.2 \sigma$ or $2 \cdot L_{cc}\cdot sin(\frac{\pi}{3})$}
\label{figure:height2}
\end{center}
\end{figure}

We continue our study for a system of film thickness $27.2\sigma$ and we observe
again the formation of semicylinders near the patterned surface and the appearance of 
defects further away. We additionally find the formation of 
cylinders perpendicular to the surface near the upper wall (Figure \ref{figure:height2}) 
in agreement with previous literature findings~\cite{wang3}.

\begin{figure}[htp]
\begin{center}
\includegraphics*[width=5.8cm,angle=90]{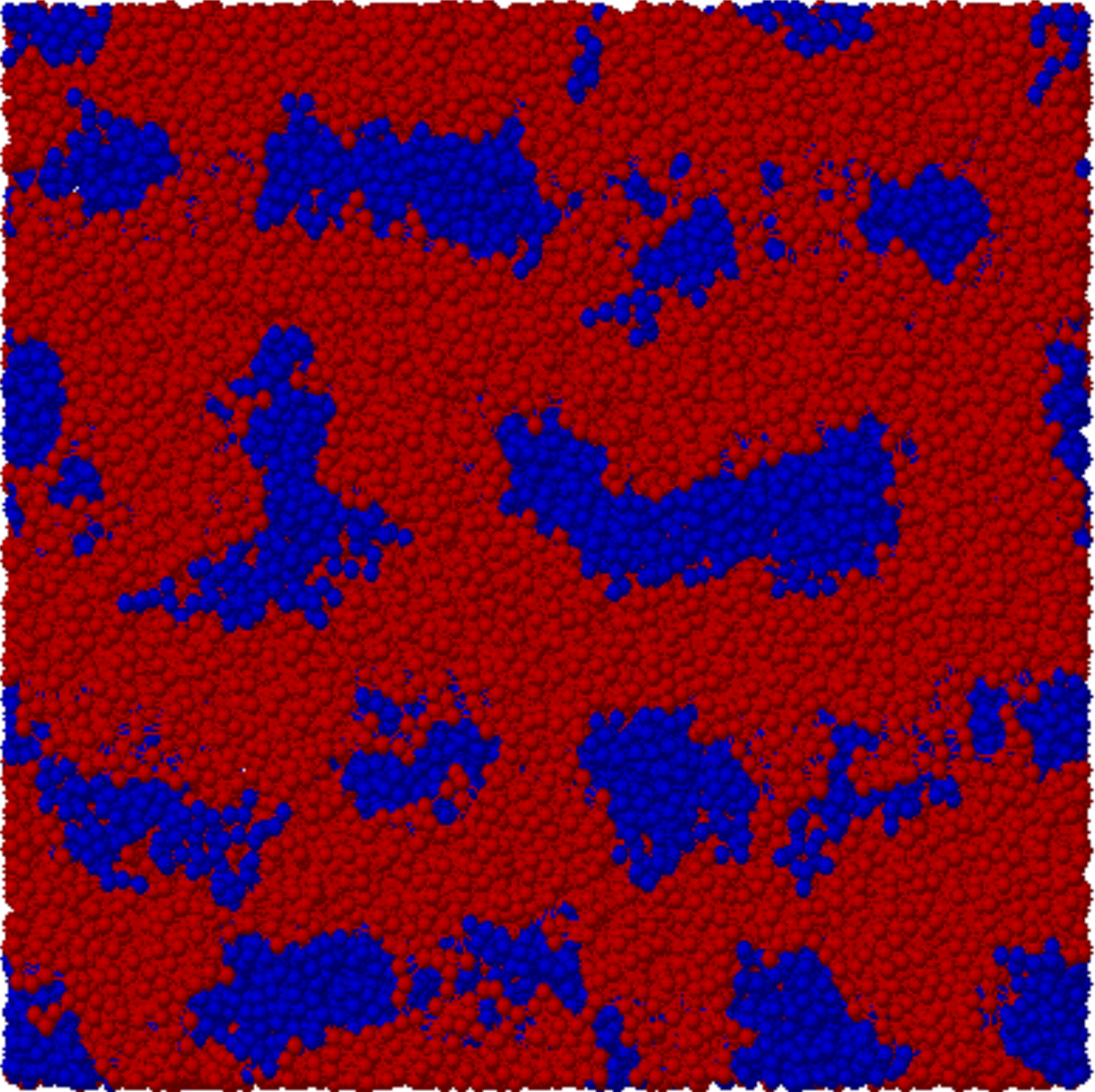}\\
\includegraphics*[width=5cm,angle=90]{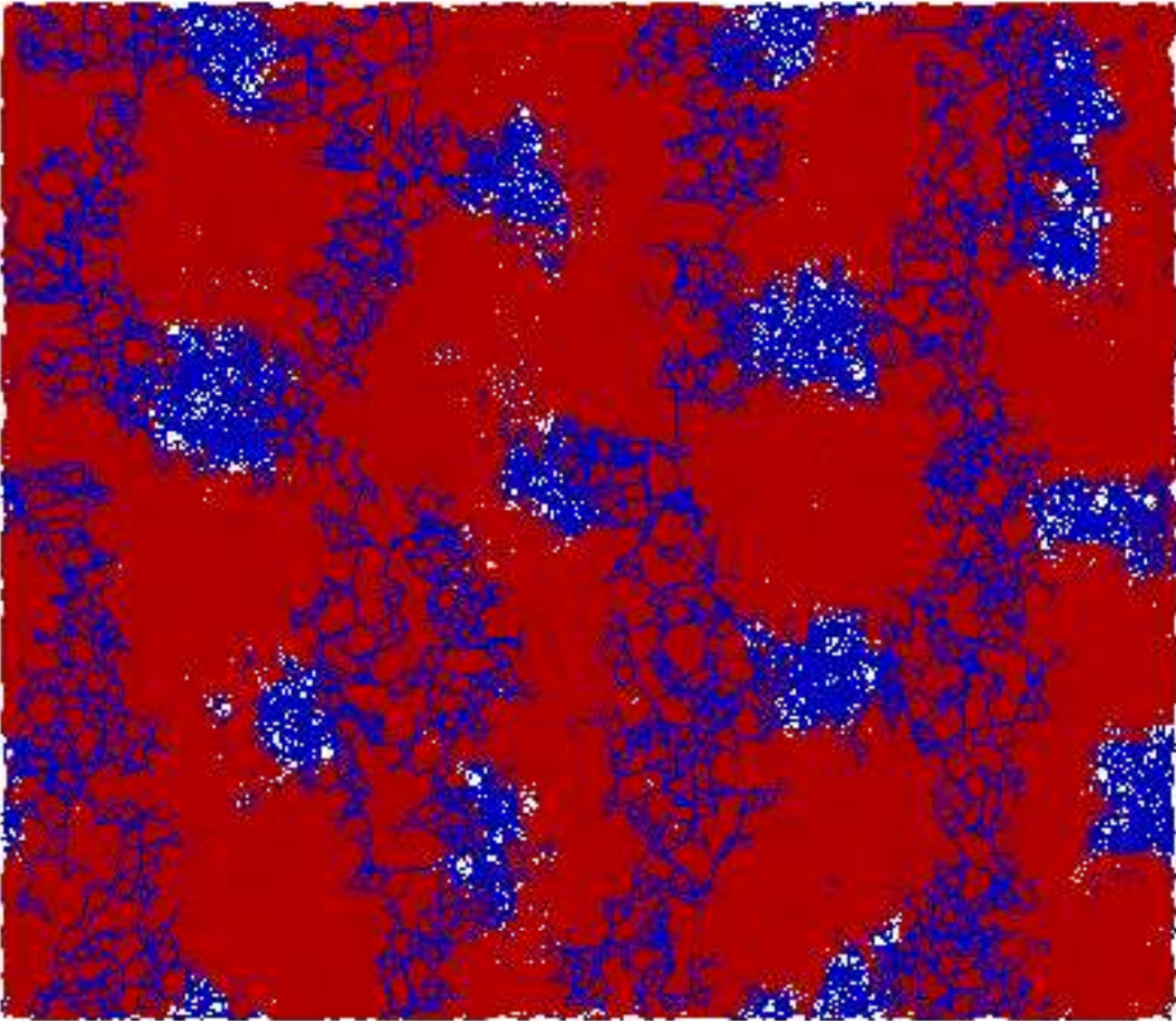}\\
\includegraphics*[width=5cm,angle=90]{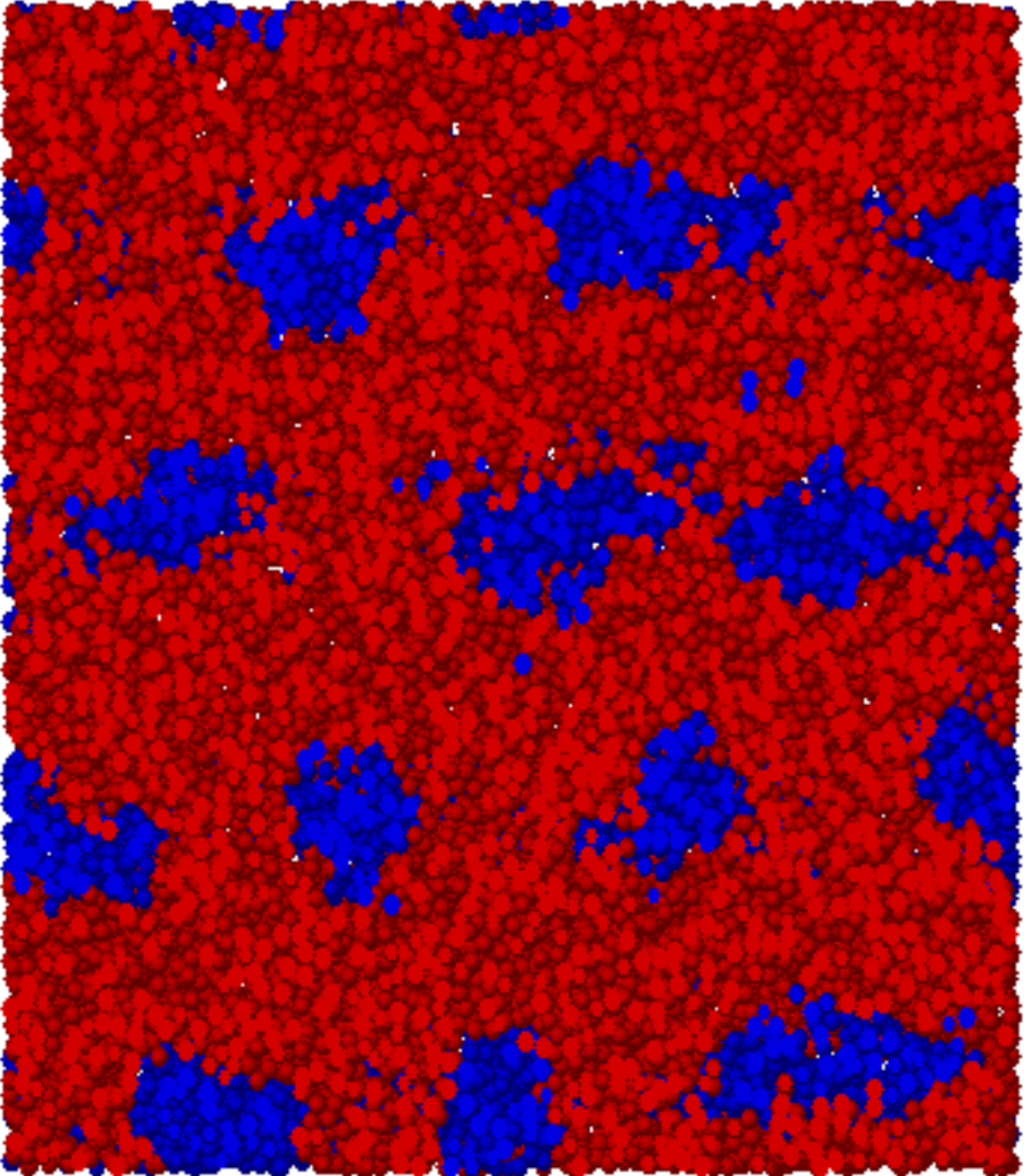}\\
\includegraphics*[width=5cm,angle=90]{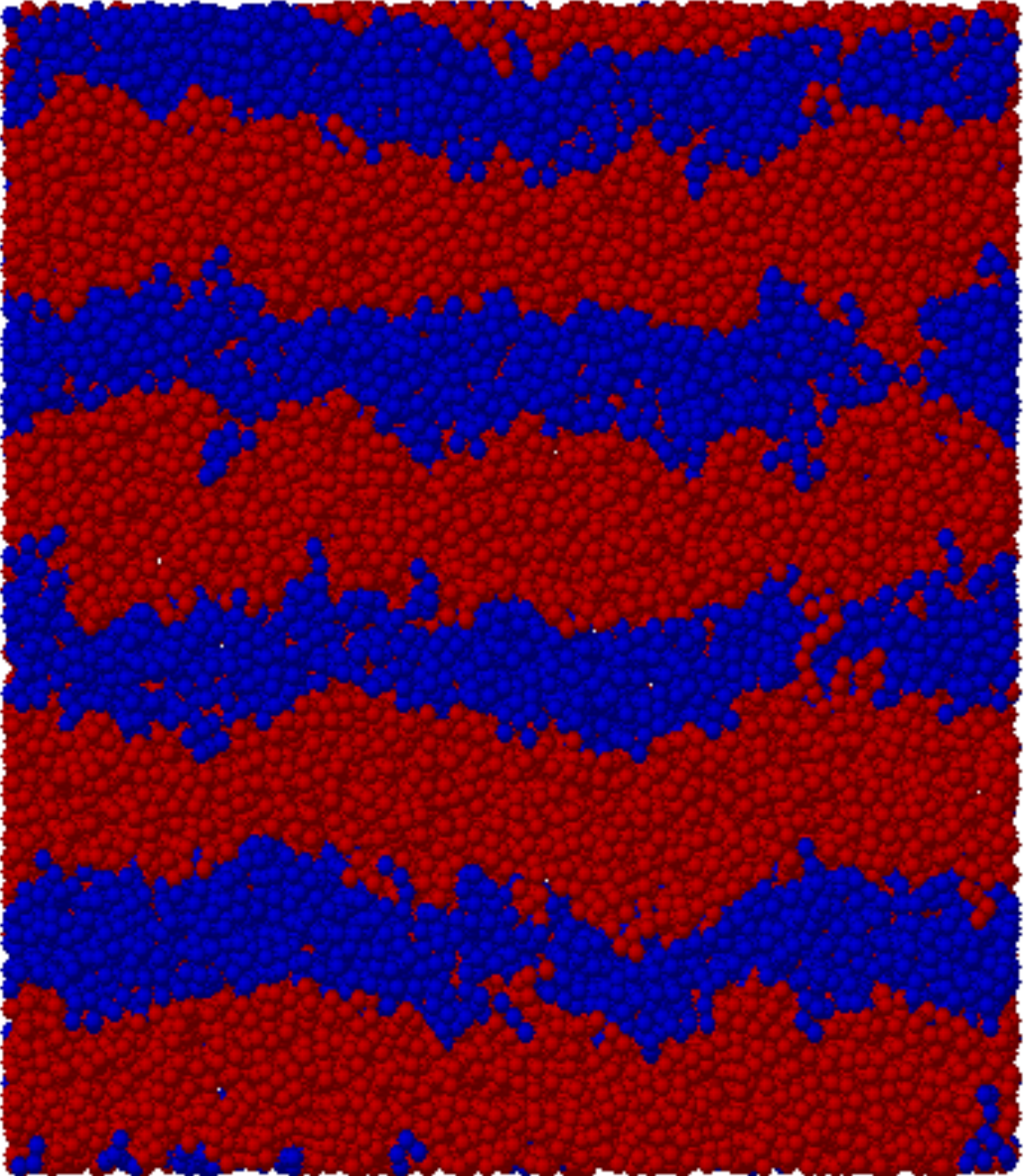}
\caption{Captions of self assembled cylindrical domains in a $13.6 \sigma$ thickness block
copolymer film. a) Top view of a system of stripe asymmetry $W = 0.55$. b) Top view of a
system of stripe asymmetry $W = 0.45$. c) Slab of the film at height $11.5 \sigma$ d) Bottom
view of a system of stripe asymmetry $W = 0.45$.
}
\label{figure:partoper}
\end{center}
\end{figure}

We saw the effect of film thickness on the self assembly of a cylindrical block copolymer 
on a patterned surface. Until now the stripe patterned surface was kept symmetric 
($W = 0.5$). By choosing a film of thickness $L_{cc}\cdot sin(\frac{\pi}{3})$, we 
perform two calculations of asymmetrical stripes one of $W = 0.45$ and a second of 
$W = 0.55$. For the system of $W = 0.55$, defects were found that consist mostly of 
unregistered cylindrical domains (Figure \ref{figure:partoper}a). More interesting 
are the results for $W = 0.45$. As we can see from Figures \ref{figure:partoper}b,c,d, 
some defects are apparent on the free surface of the film. However, if we look at a 
slab of the film slightly lower (height $11.5 \sigma$) than the free surface, we 
observe cylinders perpendicular to the surface. On the patterned surface a layer 
of semicylinders parallel to the stripes is found. These parallel semicylinders and 
the perpendicular cylinders are connected. We can see that while experiments such as 
scanning electron microscopy can provide us with a clear view of the free surface of 
the film, simulations are necessary to characterize the self assembly through the material.

\section{Conclusions}
We have described a technique, that allows us to map a coarse grained
model exactly to theory and connect it to a realistic experimental system.
For this procedure three variables are necessary to be estimated: the Flory-Huggins
parameter, $\chi N$, which is a measure of the incompatibility of different type
of segments, the end-to-end distance of the blockcopolymer chain $R_{e}$ for the
length scaling, and the invariant degree of polymerization $\overline{N}$ related
to the strength of the fluctuations. For blockcopolymer chains under confinement
one extra parameter has to be determined, which describes the strength of the polymer
segment-surface interaction controlled by $\Lambda N$.

It has to be mentioned that, most commonly for experiments and coarse grained models, 
the random phase approximation (RPA)~\cite{gennes} is used in order to obtain the 
value of $\chi$ for a system. It has been shown, though, that an alteration of the end to end 
distance is apparent with variance of interaction strength~\cite{sariban}, suggesting 
that the RPA is not accurate. In addition, our verification that $\chi$ can depend 
non-linearly on temperature suggests that RPA usage under the assumption that a 
linear temperature dependence exists can lead to incorrect results. The same is 
true for other techniques, such as the one-fluid approximation~\cite{grest}, that 
are used to obtain a value of $\chi$ in the disordered regime and then extrapolate 
the value of $\chi N$ in the ordered regime. This method will not be accurate because 
of the non-linearity of $\chi$ with temperature. The Flory Huggins lattice model is 
indeed an oversimplification. However, mapping the results on it is necessary
for a connection of experiments and simulations. The procedures that we presented 
outputs the values for the Lennard Jones parameters in order for our bead
spring model to represent a specific value of $\chi N$ in the Flory Huggins theory.

We have presented results of microphase separation of block copolymer chains
on surfaces and we found well agreement with experiments. Lamellae structures,
formed on stripe patterned surfaces, were able to accommodate mismatch between
the block copolymer lamellae period and the surface spacing. A compression or
expansion of $10\%$ gave perfect lamellae without defects. For higher
compressions or extensions, defects became apparent.

A different region of the phase diagram of the bulk block copolymer was also
investigated where by appropriately selecting the chain asymmetry and value
of Lennard Jones parameters we captured the cylinder phase. After obtaining
the hexagonal cylinder morphology, we calculated the equilibrium cylinder to
cylinder spacing. The self assembly of this system on striped patterned surfaces
of varying stripe symmetries and for different film thicknesses was studied. 

The coarse grained model described in this paper successfully captures
experimentally observed behaviors and it can be used to represent specific
realistic systems even if no atomistic details are accounted for. This
method can be used to obtain the values for the required parameters and
then one has the liberty to choose either Monte Carlo or molecular
dynamics simulations to study a problem. In our work we chose a Monte
Carlo algorithm to investigate the microphase separation. By implementing
the double bridging move we can efficiently sample the systems under
study. On the other hand, molecular dynamics can be utilized, after
the bead spring model is mapped, to investigate the dynamics of the
self-assembly.

\section{Acknowledgments}
This work is supported by NSF. Partial support from the Semiconductor Research 
Corporation (SRC) is also gratefully acknowledged.

\bibliography{nc}

\end{document}